\documentclass[11pt]{article}

\usepackage{fullpage}
\usepackage{amssymb,amsmath,amstext}

\usepackage[lofdepth,lotdepth]{subfig}
\usepackage{epsfig}
\usepackage{epstopdf}
\usepackage{graphicx}
\usepackage{pstricks}
\usepackage{algorithm,algorithmic}

\newcommand{\R}{\mathbb{R}}
\newcommand{\eqdef}{\overset{\text{def}}{=}}

\begin{document}
\title{\bf Optimal diagnostic tests for sporadic Creutzfeldt-Jakob disease based on support vector machine classification of RT-QuIC data}

\author{William Hulme$^*$  \and  Peter Richt\'{a}rik\footnote{School of Mathematics,  James Clerk Maxwell Building, King's Buildings,   Edinburgh, EH9 3JZ}	\and Lynne McGuire\footnote{National CJD Surveillance Unit,   Western General Hospital,   Edinburgh, EH4 2XU} \and Alison Green$^\dagger$}


\date{December 10, 2012}

\maketitle
\begin{abstract}

In this work we study numerical construction of optimal clinical diagnostic tests for detecting sporadic Creutzfeldt-Jakob disease (sCJD). A cerebrospinal fluid sample (CSF) from a suspected sCJD patient is subjected to a process which initiates the aggregation of a protein present only in cases of sCJD. This aggregation is indirectly observed in real-time at regular intervals, so that a longitudinal set of data is constructed that is then analysed for evidence of this aggregation. The best existing test \cite{McGuire11,McGuire11b} is based solely on the final value of this set of data, which is compared against a threshold to conclude whether or not aggregation, and thus sCJD, is present. This test criterion was decided upon by analysing data from a total of 108 sCJD and non-sCJD samples, but this was done subjectively and there is no supporting mathematical analysis declaring this criterion to be exploiting the available data optimally. This paper addresses this deficiency, seeking to validate or improve the test primarily via support vector machine (SVM) classification.  Besides this, we address a number of additional issues such as i) early stopping of the measurement process, ii) the possibility of detecting the particular type of sCJD and iii) the incorporation of additional patient data such as age, sex, disease duration and timing of CSF sampling into the construction of the test.




\textbf{Keywords:} Creutzfeld-Jakob disease, support vector machines, real-time quaking-induced conversion, RT-QuIC, coordinate descent method

\end{abstract}

\newpage

\section{Introduction}

\textbf{Background.} Current CSF tests for sCJD rely on the detection of surrogate markers of neuronal damage such as CSF 14-3-3 or tau protein. However, these markers are not specific for sCJD, thus reducing the specificity (defined below) of the test. The hallmark of sCJD is the post-translational conformational change of a normal cellular protein, PrP\textsuperscript{C}, into a disease-associated form, termed PrP\textsuperscript{Sc}, which aggregates together to form plaques within the brain.

Real-time quaking-induced-conversion (RT-QuIC) is a recently developed technique \cite{wilham10,atarashi11} that exploits the ability of PrP\textsuperscript{Sc} in brain tissue or CSF from patients with sCJD to induce a hamster recombinant PrP to change shape and aggregate over time. This aggregation is observed by adding Thioflavin T (ThT) to the reaction mixture as it binds to the aggregated PrP causing a change in the ThT emission spectrum, which may be monitored over time using fluorescence spectroscopy.  Recording these fluorescence levels during the course of the RT-QuIC thus creates a longitudinal data set, interpreted as a collection of ``curves'', and this can be used to detect the occurrence of aggregation, which manifests as an increase in fluorescence over time.

A study by scientists and clinicians at the National Creutzfeldt-Jakob Disease Research \& Surveillance Unit found that data obtained using RT-QuIC analysis could provide a test with a sensitivity and specificity of 91\% and 98\% \cite{McGuire11,McGuire11b} which compares favourably against alternative sCJD tests using surrogate marker proteins; CSF 14-3-3 (93\% and 56\%) and CSF tau protein (93\% and 79\%).

\textbf{Methods and data.} CSF samples were taken from 55 neuropathologically confirmed sCJD patients (30F:23M; aged 31-84 years; mean $\pm$ SD 62.1 $\pm$ 13.5 years) and from 53 patients (26F:27M; aged 43-84 years; mean $\pm$ SD 67.8 $\pm$ 10.4 years) who were initially suspected of having sCJD but were shown to have an alternative diagnosis. The 55 positive samples can be further categorised into three sCJD subtypes; 30 were classified as \emph{CJD-MM}, 17 \emph{CJD-MV} and 8 \emph{CJD-VV}.  The clinical details of these sCJD types are detailed in \cite{McGuire11}.

The RT-QuIC analysis was performed in quadruplicate, with fluorescent unit readings (rfu) for each of the replicates being taken every 30 minutes for 90 hours. This gives a total of $4 \times 181= 724$ readings for each sample. RFU readings are capped at 65,000 rfu.

Together with the longitudinal fluorescence data obtained, the following patient data are also available:
\begin{itemize}
\item Sex (M/F);
\item Age at first symptoms, in years;
\item LP Date (this is the date on which the CSF samples were obtained, so called due to the Lumbar Puncture procedure used to obtain the CSF);
\item Duration of disease, in months (this is the time to death from first symptoms);
\item Time to LP, in months (this is the time of first symptoms until the time the CSF sample was collected).
\end{itemize}
Information on the final two factors is unavailable for non-sCJD cases.

\textbf{The test.} By subjectively examining the plotted rfu data over time, a novel sCJD specific test using the RT-QuIC technique was proposed in \cite{McGuire11,McGuire11b}, where a positive result was defined as the mean of the two highest rfu readings at 90 hours being over 10,000 rfu. It is this test with this single criterion which provides the high sensitivity and specificity values above. This test criterion essentially discards all but the final rfu reading, and it does not use any of the additional data listed above.

\textbf{Problem statement.} There are no rigorous mathematical analyses that suggest this test exploits the rfu data optimally, nor that it will reliably generalise to new CSF samples whose sCJD status is unknown, as is its purpose. It is demonstrably superior to CSF 14-3-3 and CSF tau protein tests, but it may be possible to increase the sensitivity and specificity further by using more and possibly all of the available data. Further, \cite{McGuire11,McGuire11b} does not consider the potential of the rfu data to detect differences \emph{between} sCJD types. These issues are addressed in this paper.

\subsection{Preliminary analysis}

\begin{figure}[h!]
\centering
\includegraphics[width=0.9\linewidth]{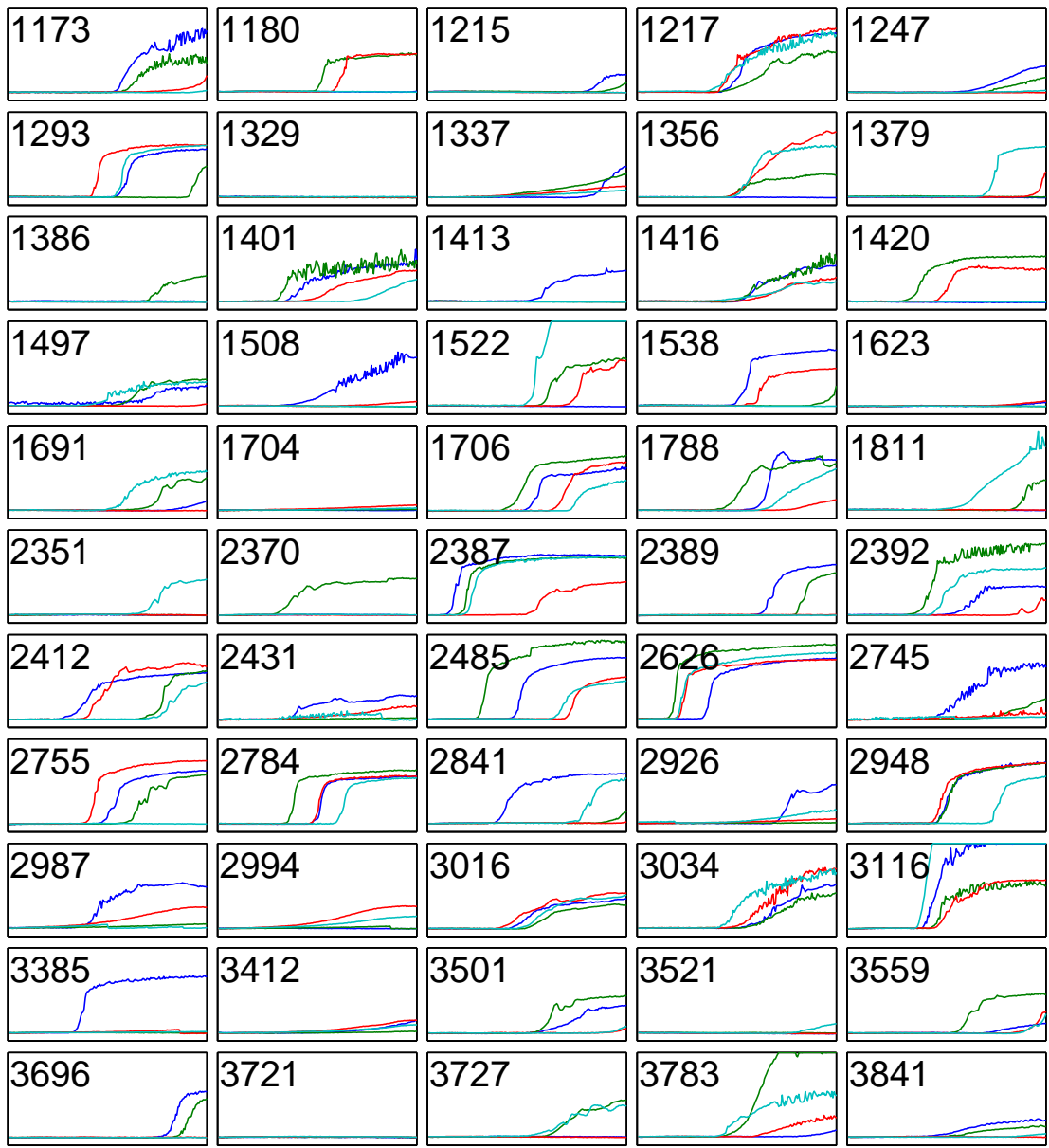}
\caption{Plots of quadruplicates for samples classified as being sCJD positive. The horizontal axis runs from 0 to 90 hours, and the vertical axis runs from 0 to 65,000 rfu.}
\label{quad1}
\end{figure}

\begin{figure}[h!]
\centering
\includegraphics[width=0.9\linewidth]{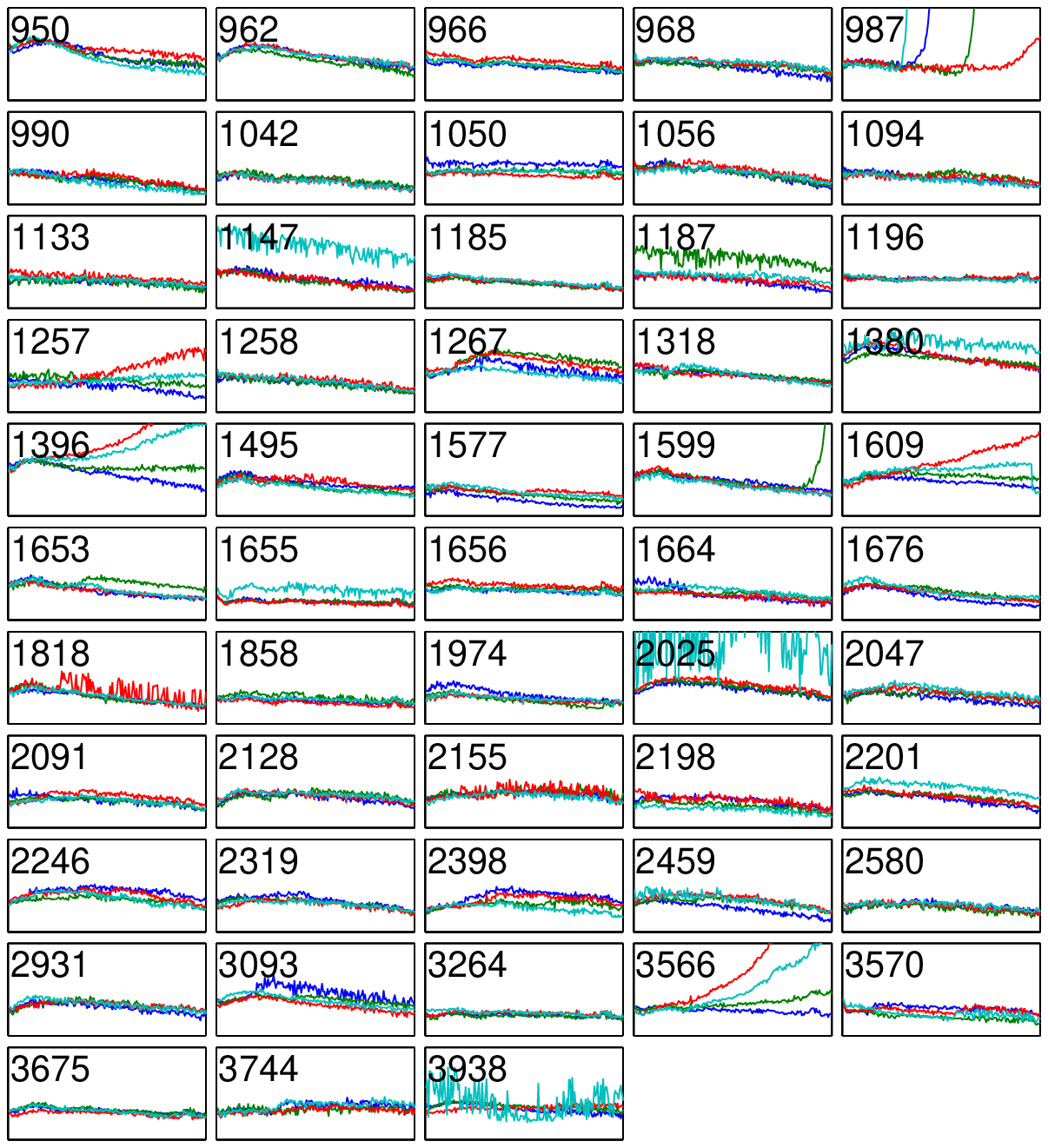}
\caption{Plots of quadruplicates for samples classified as being sCJD negative. The horizontal axis runs from 0 to 90 hours, and the vertical axis runs from 4500 to 8000 rfu.}
\label{quad-1}
\end{figure}

The graphs in Figures~\ref{quad1} and \ref{quad-1}, which plot the fluorescence readings over time for all samples and replicates, show obvious differences between sCJD positive and sCJD negative cases, henceforth simply \emph{positive} and \emph{negative}.  Note that the scale for positive cases is much larger than the scale for negative cases.  To get an idea of the relative difference between these two classes, see Figure~\ref{rawdataplot}.   As expected, the \emph{curves} for the positive cases in general exhibit an increase in rfu over time whereas the negative curves tend to remain roughly constant. There are some exceptions, most notably sample \#987, which despite being classified as negative, shows an increase in rfu beyond doubt in 3 of its replicates. The maximum rfu reached for each of these three curves ranges from 34,212 to 42,145 rfu, but this is hidden by the range of the vertical axis. \cite{McGuire11,McGuire11b} explains that although this case was classified as negative, sCJD could not be ruled out entirely.

Noise levels, that is, the random discrepancies between the idealised rfu level over time and that which is observed, can vary greatly between and within samples. Some curves display high levels of noise whereas others appear much smoother. Discussions with the authors of \cite{McGuire11,McGuire11b} explain that there are mechanical explanations for noise that exceeds the usual levels, relating to equipment faults. The noise level is not thought to relate to the sCJD status of the patient.

The general profile of each of the four curves from a given sample can differ greatly. Most positive cases are characterised by three distinct states over the 90 hours:
\begin{itemize}
\item an initial base-line period of a constant low-level rfu (5,000-6,000 rfu);
\item an increase in rfu which may be anywhere between extremely rapid (up to the maximum fluorescence within hours) to very slow (appears to still be increasing at 90 hours by which time rfu has only doubled from the base-line value);
\item a final period where the rfu is constant or slowly increasing.
\end{itemize}
The times at which the curve passes from one state to the next varies greatly, as does the final rfu reached, even within samples. Some positive samples only have one curve displaying any clear evidence of aggregation while the remaining curves have a constant, low-level rfu over time.

Clearly, there is a great deal of variation between and within samples and the most meaningful way to use the data from the four replicates to diagnose a sample is not immediate. It is unknown whether any of this variation is due to factors such as age, sex or disease duration. The only indisputable source of variation is the presence or not of sCJD in the sample and this is enough to provide a high-performance test for sCJD as detailed in \cite{McGuire11,McGuire11b}, henceforth the \emph{null test}. However, it is conceivable that within all the data available, enough information is present to make a better or possibly perfect distinction between the two classes, that generalises well to new observations.

Many alternate and potentially better tests might be proposed using educated guesses and trial and error, but a systematic approach---that can automatically find the important features of the data, and use it optimally in some sense---would be useful here, along with well-defined criteria to adequately compare tests.

There is a machine learning / mathematical programming  technique known as \emph{support vector machine} classification \cite{Bishop06} which can be employed to approach the problem in this way, and it is this technique that will be used throughout this paper.

\section{Support vector machine classification}\label{chap:SVMs}
Primarily attributed to the work of Vladimir Vapnik \cite{Vapnik82,Vapnik96,Vapnik98}, support vector machines (SVMs) are a powerful and widely-applicable machine learning technique used for binary classification of data. Applications include handwritten digit recognition \cite{Vapnik95}, text categorisation \cite{Joachim} and bio-informatics \cite{Ivaniuc07}. SVMs operate by representing the data as points in a multi-dimensional space and finding a hyperplane which  separates the two classes, if separation is possible. Certain separating hyperplanes are preferred to others. Ideally, we would wish the nearest point in either class from the hyperplane to be as far a possible; this concept is referred to as \emph{margin maximisation}. The classification of a new observation is performed according to which side of the gap it falls.

The arguments that motivate the SVM formulation and present its properties draw from a range of mathematical ideas, and are an accumulation of decades of progress in machine learning theory. Presented here is a succinct summary of these arguments (which draws upon the expositions in \cite{Bishop06,Hastie09,Cristianini00}), and a discussion of two classification-improving SVM methods that are heavily used throughout the paper.

\subsection{Training set and separating hyperplane}

Consider a set of $n$ multivariate observations/examples, $x_i\in\mathbb{R}^m$, $i=1,\dots,n$, where each observation is one of two types/classes. The class of observation $x_i$ is denoted by $y_i$; we assume it is \emph{known} for all $i$. We thus have $n$ pairs $(x_i,y_i)$ forming the \emph{training set}
\begin{equation*}
T \eqdef \{(x_1,y_1),(x_2,y_2),\dots,(x_n,y_n) \}.
\end{equation*}
The training set $T$ is  used to construct/train a model which is then used to classify new observations the class of which is  \emph{not known}.

A typical application is that of spam filtering, where the vectors $x_i$ correspond to emails and the label $y_i$ identifies each email as spam or not spam. In this paper the observation $x_i$ corresponds to longitudinal data corresponding to CSF sample of patient $i$, and $y_i$ identifies each patient as having  CJD or not.

An approach to classification, common in the machine learning and mathematical optimization literature,  is to seek a hyperplane in $\mathbb{R}^m$ that separates the two classes of observations contained in the training set, if possible.  Formally, given a training set $T$, we wish to compute
a \emph{weight vector}  $w\in \mathbb{R}^m$ and \emph{bias} $b\in \mathbb{R}$, so that
\begin{equation}\label{constraint}
y_i(w^t x_i - b) \geq 0, \qquad i=1,\dots,n,
\end{equation}
where $u^t v$ denotes the standard inner product between vectors $u,v \in \R^m$: $u^t v = \sum_{i} u_i v_i$. Note that, given $(w,b)$ satisfying \eqref{constraint}, the positive and negative observations lie in different half-spaces generated by the hyperplane
\begin{equation*}
(w,b) \eqdef \{x \in \mathbb{R}^m \;:\; w^t x - b = 0\}.
\end{equation*}
Indeed, observations  $x_i$ with $y_i=+1$ lie in the half-space $\{x\in \R^m \;:\; w^t x-b \geq 0\}$, and observations $x_i$ with $y_i=-1$ lie in the half-space $\{x\in \R^m \;:\; w^tx-b \leq 0\}$.

The class $y$ of a \emph{new observation} $x$  is estimated/determined by finding which ``side'' of the hyperplane this point lies on, i.e., by setting
\begin{equation*}
y \eqdef \text{sign} \{w^t x - b\}.
\end{equation*}
By convention, we may set $\text{sign}\{0\} = 1$ (or break the tie arbitrarily).

\subsection{The maximum margin hyperplane}

If a hyperplane $(w,b)$ satisfying (\ref{constraint}) exists, we say that the training set $T$ is \emph{linearly separable}. In such a case, usually there is an infinity of separating hyperplanes. There are several algorithms which can be used to identify/compute one of these hyperplanes (see \cite{Hastie09}, Chapter 4). The SVM approach to this problem is to choose a hyperplane maximising the distance (so called \emph{margin}) to the nearest training point $x_i$. The distance from $x_i$ to the hyperplane $(w,b)$ is given by
\begin{eqnarray}\delta_i &\eqdef& \frac{|w^t x_i + b|}{\|w\|_2},\end{eqnarray}
where $\|w\|_2 \eqdef (\sum_i w_i^2)^{1/2}$ is the $L_2$ norm of $w$. If $(w,b)$ separates the two classes, then $|w^t x_i + b| = y_i(w^t x_i +b)$ for all $i$, and we can find the maximum-margin hyperplane via solving the  following optimisation problem:
\begin{equation}\label{SVM:margin}
\underset{w, b}{\text{maximise}}   \min_{i=1,2,\dots,n} \left\{\frac{y_i(w^t x_i + b)}{\|w\|_2} \right\}.
\end{equation}

In this form, the optimization problem \eqref{SVM:margin} is not easily solvable. However, noting that the objective function in \eqref{SVM:margin} is positively homogeneous of degree one (i.e., $\delta_i$ does not change if we replace $(w,b)$ by $(tw,tb)$ for $t>0$), we may further  reformulate \eqref{SVM:margin} as
\begin{equation}\label{SVMcanonical}
\begin{aligned}
& \underset{w, b}{\text{minimise}} & & \tfrac{1}{2}\|w\|_2^2 \\
& \text{subject to } 						& &  y_i(w^t x_i - b) \geq 1, & i=1,2,\dots,n,
\end{aligned}
\end{equation}
which is a quadratic minimisation problem with $n$ linear constraints. This problem can be solved efficiently using standard convex optimization algorithms \cite{Ben-Tal_Nemirovskii:2001:book} such as interior point methods \cite{Nesterov-Nemirovski:1994:IPMbible,Renegar:1991:book}.

Let $(w,b)$ be the optimal solution of \eqref{SVMcanonical}. Points $x_i$ for which the constraint $y_i(w^t x_i - b) \geq 1$ is active, i.e., for which it holds as an equality, are referred to as \emph{support vectors}. Note that it is precisely the support vectors which define the hyperplane. Indeed, any non-support vectors can be removed from the training set and the solution $(w,b)$ will not change. The act of solving the optimisation problem \eqref{SVMcanonical} is usually called \emph{training a support vector machine.}

\subsection{Soft margins}\label{subs:soft}
In general, linear separability of the training set may not be possible in which case the optimization problem \eqref{SVMcanonical} is \emph{infeasible} (i.e., there is no pair $(w,b)$ satisfying the $n$ constraints) and hence cannot be used to find $(w,b)$.

One approach to this  is to map the data into a space in which separation is possible (so called \emph{kernel trick}); this is not considered in this paper. An alternative approach, pioneered by Vapnik \cite{Vapnik95}, is to allow misclassifications but to  discourage this by penalization. We will now describe this approach.


For each observation $(x_i,y_i)$ in the training set $T$ we introduce a slack variable $\xi_i\geq 0$ measuring the degree to which the $i$-th inequality/constraint in  \eqref{SVMcanonical} is not met. We then relax the $i$-th inequality to $y_i(w^T x_i + b) + \xi_i \geq 1$, and add the term $C\sum_i \xi_i$ to the objective function, where $C\geq 0$, which has the purpose of pushing the slack variables to zero. This leads to the following optimisation problem:
\begin{equation}\label{SVMsoftmargin}
\begin{aligned}
& \underset{w, b, \xi}{\text{minimise}} & & \tfrac{1}{2}\|w\|_2^2 + C\sum_{i=1}^n \xi_i \\
& \text{subject to } 						& &  y_i(w^t x_i - b) + \xi_i \geq 1, & i=1,2,\dots,n,\\
& 													& & \xi_i \geq 0, & i=1,2,\dots,n.
\end{aligned}
\end{equation}
Note that if $0\leq \xi_i < 1$, $x_i$ is classified correctly. If $\xi_i=1$, $x_i$ may or may not be classified correctly. If $\xi_i>1$, $x_i$ is misclassified. Whenever $\xi_i > 0 $, penalty $C\xi_i$ is incurred.

As $C$ increases, the optimal hyperplane will misclassify fewer observations, and for large enough $C$ the hyperplane will fit as tightly as possible to minimise the sum of the slacks. Once this state has been reached, increasing $C$ further has no effect on the solution, other than to increase optimal objective function value.

The hard-margin case in (\ref{SVMcanonical}) corresponds to $C=\infty$. In this case, if the training set is not linearly separable, then the objective function value of (\ref{SVMsoftmargin}) is infinite because at least one $\xi_i$ must be non-zero, rendering the solution infeasible, and this agrees with the infeasibility of (\ref{SVMcanonical}) for non-separable problems. If the problem is separable, then the $\xi_i$s will be forced to zero by the minimisation, thus removing the slack variables and recovering the original problem.

\subsection{Feature Selection}\label{subs:sparsity}
Maximisation of the margin between the two classes may not be the only criterion for obtaining a good classifier. Sometimes it is desirable that the dimension $m$ of the space in which the observations are expressed is reduced so that only those variables, or \emph{features}, which contribute the most to the separability of the training set are used, so that the unnecessary features do not over-complicate the model. Reducing the number of non-critical features expressed in $w$ may be desirable for easier interpretation of the classifier. This process is known as \emph{feature selection}.

A widely used technique for encouraging $w$ to be sparse is to introduce the sparsity-inducing term term $D\|w\|_1$ into the objective function of the minimisation problem \cite{Tibshirani96,Zou05}, where $D\geq 0$ is a constant and $\|w\|_{1}=\sum_i |w_i|$ is the $L_1$ norm of $w$. Larger values of $D$ encourage more sparsity in $w$. Thus, instead of \eqref{SVMsoftmargin} we are interested in solving the optimization problem
\begin{equation}\label{SVMcomplete}
\begin{aligned}
& \underset{w, b, \xi}{\text{minimise}} & & \tfrac{1}{2}\|w\|_2^2 + C\sum_{i=1}^n \xi_i + D\|w\|_{1} \\
& \text{subject to } 						& &  y_i(w^t x_i - b) + \xi_i \geq 1, & i=1,2,\dots,n,\\
& 													& & \xi_i \geq 0, & i=1,2,\dots,n.
\end{aligned}
\end{equation}
%

Clearly, \eqref{SVMsoftmargin} is recovered for $D=0$ and \eqref{SVMcanonical} is recovered for $D=C=0$. By increasing $D$, the number of zeros in the optimal vector $w$ will grow. More interestingly, by proper choice of $C$ and $D$ a balance can be struck between the conflicting goals of finding $w$ supported at a few  of the most important features only (controlled by $D$) and seeking an acceptably large enough margin (controlled by $C$).


Huge-scale $L_1$-regularized optimization problems can be efficiently solved by coordinate descent methods \cite{RT2012,RT2011,RT-TTD-2012}. Greedy coordinate descent methods for $L_1$-regularized problems were first analyzed in \cite{RT-TTD-2012}, a general theory of serial and parallel randomized methods is developed in \cite{RT2011,RT2012}.



\section{Performance measures}
Before applying a SVM classifier to new data, performance measures need to be devised for easy comparison between alternate tests/approaches. Sensitivity (true positives / total positives) and specificity (true negatives / total negatives) have already been mentioned, and these give a good, quantitative account of how many observations in each class are correctly separated by the hyperplane of any given classifier. These measures together will be referred to as the \emph{fit} of the test. Optimising the fit results in a test which minimises the number of classification errors on observations from the training set.

An overly good fit on the training data may not necessarily lead to  reliable classifications on \emph{new data}. Indeed, the SVM may be \emph{overfitting} to the training data, increasing generalization error (which we refer in this paper by the term robustness). \emph{Robustness} may be difficult to quantify if there is no new data with which to assess the test. A work-around is to train a SVM on  a portion of the data only and then assess the performance of the resulting test on the remaining data. This is called \emph{cross-validation} (CV) \cite{Hastie09} and can provide useful feedback regarding the amount of overfitting afflicting a given test.

The particular CV technique proposed here is \emph{leave-one-out}, which for a given training set $T$ proceeds as follows: for each observation $x_i$, train a SVM on the training set from which $(x_i,y_i)$ has been removed and classify $x_i$ according to this new SVM. That is, for all $i$, compute
\begin{equation*}
y_i^\prime \eqdef \text{sign} \{(w^i)^{T} x_i - b^i\},
\end{equation*}
where $(w^i,b^i)$ is the hyperplane obtained when training an SVM on $T\backslash \{(x_i,y_i)\}$. If $y_i^\prime = y_i$ then the classification is correct, otherwise the classifier has failed for this observation.

The benefit of removing just a single observation from the training set each time is that information analogous to the sensitivity and specificity of a test can be gained by identifying the number of false positives and false negatives produced by the cross-validation. As the classifier may change for each $x_i$ we do not have performance information for a single classifier, but rather a family of classifiers created from the choice of training set and the choice of parameters $C$ and $D$. The performance information gathered in this way shall be referred to as \emph{pseudo-sensitivity} and \emph{pseudo-specificity}, and these will be the measures of robustness. A few remarks:
\begin{itemize}
\item If a training set is linearly separable for parameters $C$, $D$, then sensitivity = specificity = 100\%. However, the classifier may overfit to the data, and so the test may not be very robust.
\item If $x_i$ is not a support vector, then the classification of this vector by the cross-validation method above will be correct, because removing this vector will not alter the hyperplane. The number of support vectors in a training set is therefore an upper-bound on the number of misclassifications possible by cross-validation. Further, the number of positive support vectors bounds the number of false negatives, and the number of negative support vectors bound the number of false positives.  This provides an intuitive perspective for understanding how the number of support vectors acts as a proxy for the robustness of the classifier.
\item If a test has 100\% pseudo-sensitivity and 100\% pseudo-specificity, this is no guarantee that new observations will classified correctly.
\item  The sensitivity is an upper bound on the pseudo-sensitivity, and the specificity is an upper bound on the pseudo-sensitivity. Consequently, if the sensitivity and specificity are equal to the pseudo-sensitivity and pseudo-specificity respectively, then test can be considered to be optimally robust.
\end{itemize}

For conciseness and easy comparison, the fit of a test will be reported as the non-simplified fraction of the sensitivity and specificity values in the format \[\left(\frac{true\;positives}{total\;positives},\frac{true\;negatives}{total\;negatives}\right)_{F},\] as all else is already known and easily recovered. The robustness will be referred to in a similar format, only now the numerators are the number of positive or negative observations that, when removed from a training set, are correctly classified by the SVM trained on the reduced training set. The subscript will be $R$.

\section{SVM analysis on untransformed fluorescence data}

The first application of support vector machines in this paper is an application directly to all or a subset of the raw fluorescence data in an attempt to find superior tests to the null test. The most conspicuous way to do this is by finding a SVM-optimal alternative to the 10,000 threshold, by training a support vector machine on the training set employed by the null test. Also considered is using the single highest fluorescence reading at 90 hours from each sample as a training set.

As readings are taken every half hour it is possible to observe whether an equally high performance test is achievable \emph{before} 90 hours, so that the result of the test may be known sooner. This is investigated here by training a SVM on each of the 181 readings and observing their performance. This is first done by taking the maximum of the four readings at each time, then by training on the average of the two highest readings.

\subsection{Tests on the fluorescence reading at 90 hours}\label{subs:adjust}
Recall that the null test misclassifies only 6 from 108 observations, with a fit of $(\frac{50}{55},\frac{52}{53})_{F}$.
The threshold of this test, 10,000 fluorescent units, was decided subjectively and it can therefore be adjusted via SVMs so as to maximise the margin between the two classes. Training a soft margin SVM on this one-dimensional training set (separability is clearly not possible), an optimal bias and weight vector of $b^*=2.3089$ and $w^*=2.1580e^{-4}$ respectively are obtained, along with a maximum margin of $\frac{1}{\|w^*\|_2}=\frac{1}{w^*}=4,634$.
Recall that for a given test defined by the hyperplane $(w,b)$, the test-classification (as opposed to the true classification) of an observation $x_{i}$ is calculated by
\begin{equation}
y_{i}=\text{sign}\{w x_i - b \},
\end{equation}
when $m=1$, and from this
\begin{equation}
\begin{aligned}
x_i \geq \frac{b}{w} & \quad \Longrightarrow \quad &  y_i = +1,\\
x_i < \frac{b}{w} & \quad \Longrightarrow \quad & y_i = -1 &,
\end{aligned}
\end{equation}
can be recovered. Thus, $\frac{b}{w}$ is the threshold above which a sample is defined as positive and for the above test, the threshold is $\frac{b^*}{w^*}= 10,699$, which shows that the 10,000 threshold from the original test was not too far from the SVM-optimal threshold. No values in the training set lie between 10,000 and 10,699 so the sensitivity and specificity remain the same.

Rather than the two highest readings at 90 hours, an alternate test could be to take just the single highest reading at 90 hours. 
Training a soft margin SVM on this training set delivers an optimal bias of $b^*=1.8878$ and a maximum margin of $\frac{1}{w^*}=7,286$, which together define the threshold as 13,755 fluorescent units. The performance is the same as the null test.

As these adjusted tests are both calibrated by training a SVM, optimality with respect to the training set is guaranteed. Both tests have the same performance, but as the latter of these has a greater margin, the training set it operates on can be considered to exhibit a greater distinction between the two classes, which suggests greater classification accuracy in the face of new samples. Taking the single highest reading at 90 hours as opposed to taking the average of the two highest readings thus may be a more reliable summariser of the data.

Cross-validation shows that the robustness of both these tests matches the fit; that is, no observation is classified differently when training a SVM on a training set with any single observation removed. As such, the tests are optimally robust, strongly indicating the test will perform well when classifying new observations.

\subsection{Tests on single fluorescence readings before 90 hours}\label{90toolong}

\begin{figure}[h!]
\centering
\subfloat[including \#987]{\includegraphics[width=0.4\linewidth]{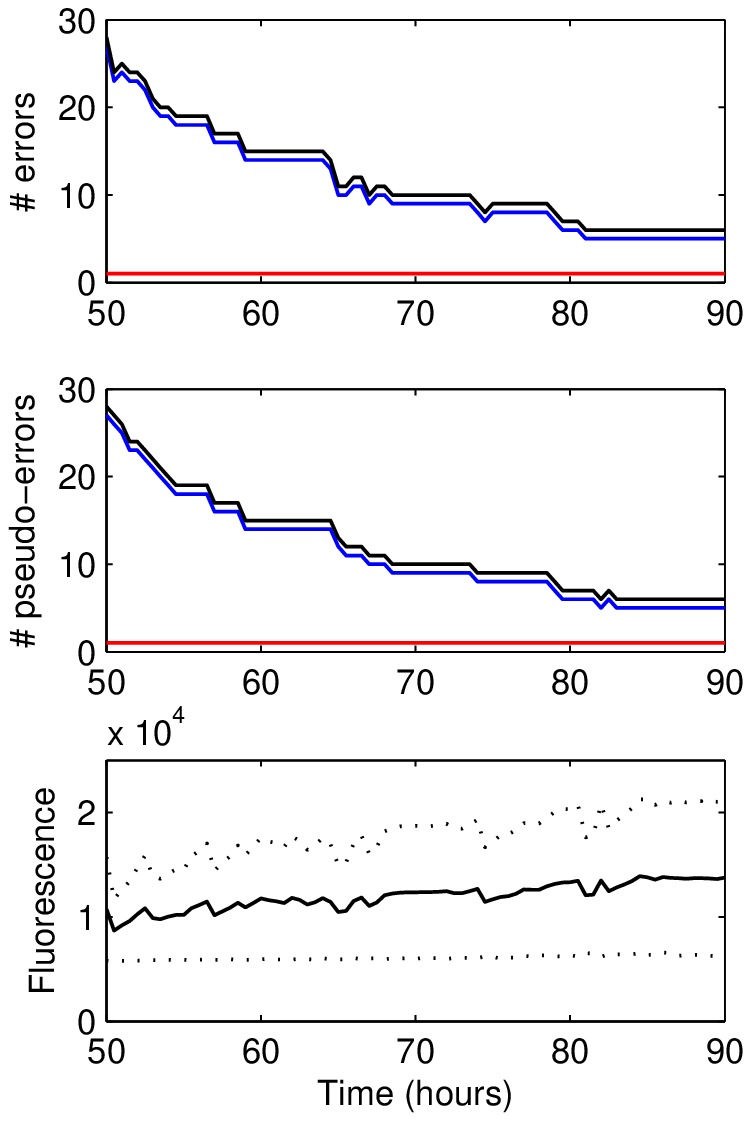}}
\subfloat[excluding \#987]{\includegraphics[width=0.4\linewidth]{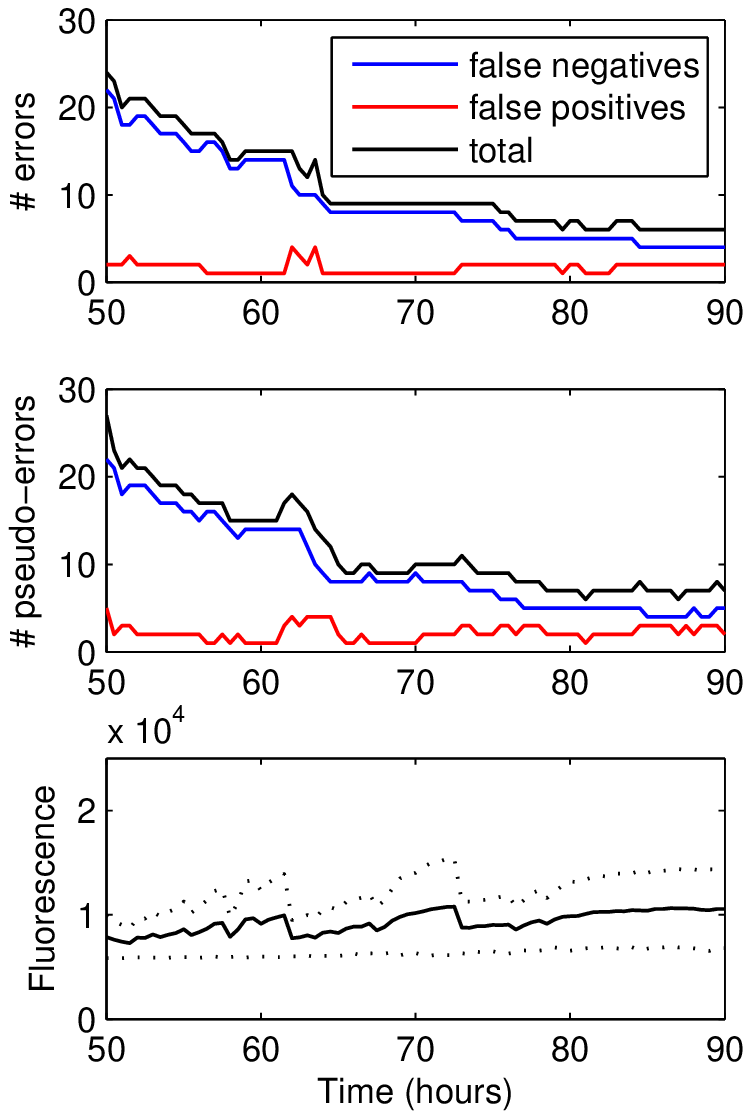}}
\caption{Performance measures of SVMs trained on the single maximum fluorescence readings for each hour from 50 to 90. The bottom panels show the threshold (solid line) together with the size of the margin either side (dotted-line). It is important to note that the progression of the threshold and margin in these graphs is the result of a succession of individual SVMs trained on single readings, and not a representation of one threshold trained on the whole data.}
\label{singletime}
\end{figure}

The approach in the preceding section is applied to every single fluorescence reading. For the single highest fluorescence reading, soft margin SVMs are used (separability is not possible) and the number of classification errors and pseudo-classification errors are recorded in Figure~\ref{singletime}(a), together with the value of the threshold and the size of margin. Measurements before 50 hours are discarded as the specificity and sensitivity is extremely poor at these times.

It can be seen that the performance steadily increases for later readings and then stabilises beyond about 83 hours. The threshold stabilises at the same time, settling at just under 14,000 fluorescent units, with a final margin of 7,286 (these are the exact same measures from the SVM taking the maximum fluorescence at 90 hours in Section~\ref{subs:adjust}, because the data is identical). Perhaps more interestingly, there is always only 1 false positive, and this corresponds to sample \#987. As discussed, sCJD could not be entirely ruled out for this sample, and given the results of the RT-QuIC analysis it is highly plausible that this sample is in fact sCJD positive. It may be interesting then to observe the performance of the SVMs trained on single readings when this sample is excluded and these results are presented in Figure~\ref{singletime}(b).

A similar pattern emerges, only this time the final threshold stabilises just above 10,000 fluorescent units, and the final margin is a much smaller 3,760. The number of false positives has actually increased overall, due to the lowering of the threshold towards the maximum readings of the negative cases. Clearly, sample \#987 is having a significant impact on the threshold, raising it by almost 4,000 fluorescent units and increasing the margin in an attempt to reduce the size of $\xi_{987}$, which contributes to the objective function.

Also interesting is that the number of pseudo-errors increases when \#987 is removed. This is likely to be caused by the stability that its inclusion causes, due to it dominating the objective function and therefore largely dictating the location of the threshold. During cross-validation, when individual samples are removed and the SVM re-trained, \#987 still exerts a big influence on the threshold, so each sub-threshold within each cross-validation loop will not change very much, and the classification is more likely to be the same as it was for the parent SVM. Remove \#987 and perform cross validation, and there is now no single sample which dominates the objective function, and the sub-SVMs have more freedom in finding the optimal threshold. The threshold will then vary more greatly and so classification of the single removed sample is less likely to match with the parent SVM. Of course, the threshold will only be altered when support vectors are removed, which is why the difference in the number of pseudo-errors between the two training sets is small. So, the inclusion of sample \#987 improves robustness, but this is due to the way in which it restricts the threshold under cross-validation, distorting the results to accommodate it.

\begin{figure}[h!]
\centering
\subfloat[including \#987]{\includegraphics[width=0.4\linewidth]{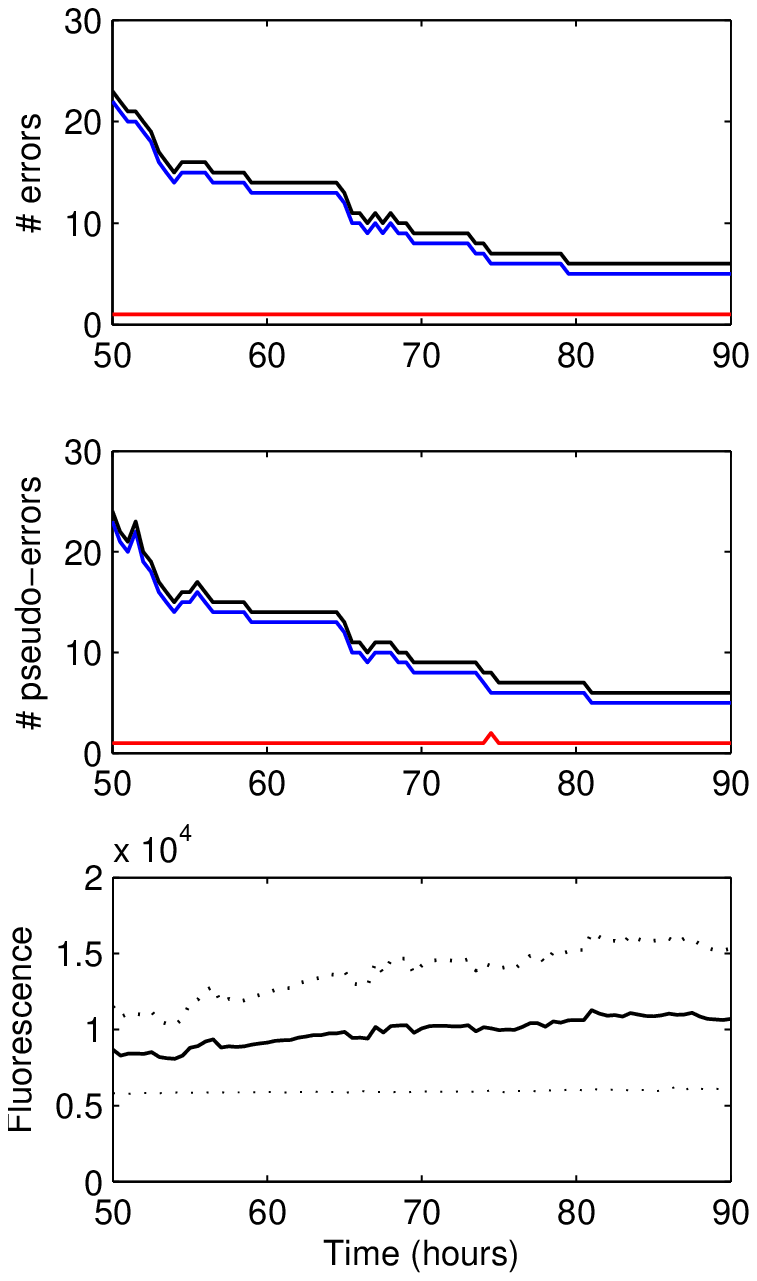}}
\subfloat[excluding \#987]{\includegraphics[width=0.4\linewidth]{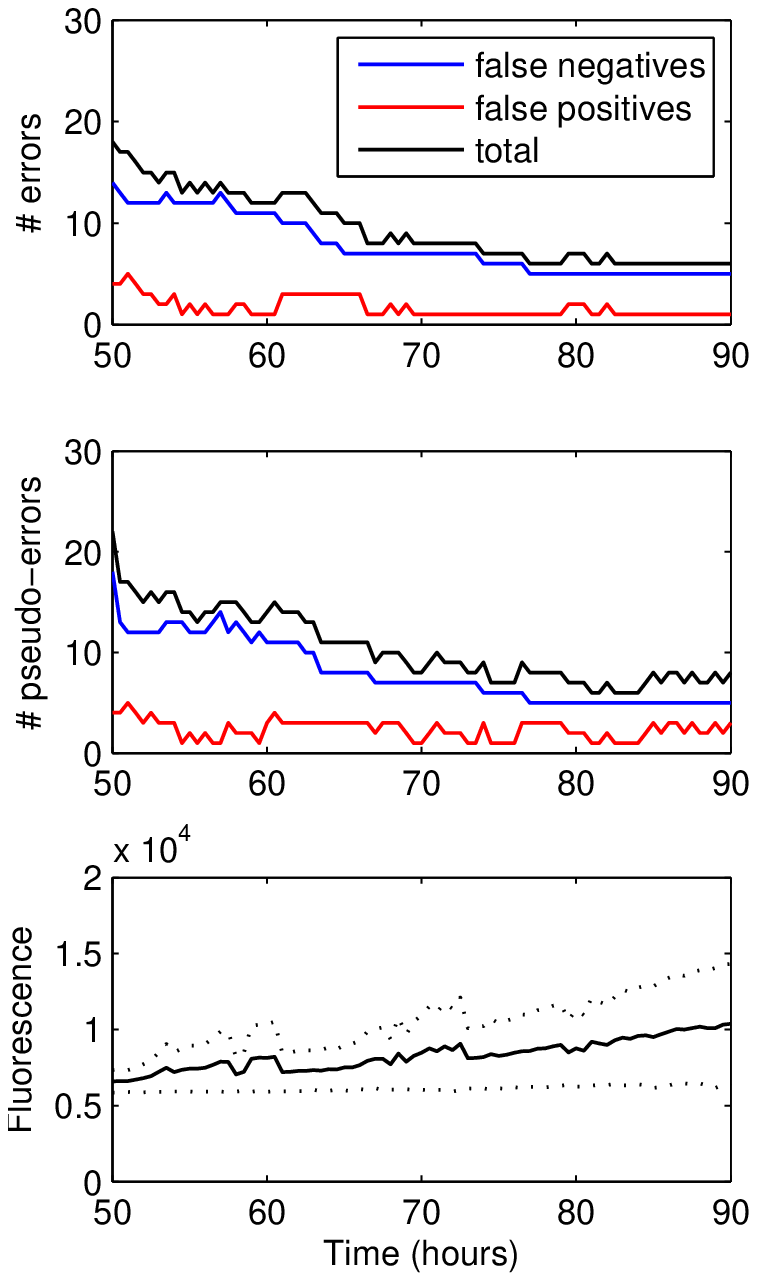}}
\caption{Performance measures of SVMs trained on the average of the two highest fluorescence readings for each hour from 50 to 90.  The bottom panels show the threshold (solid line) together with the size of the margin either side (dotted-line).}
\label{singletimeavg}
\end{figure}

The same analysis is now applied to the average of the two highest readings. The performance measures are presented in Figure~\ref{singletimeavg}, and the results are similar to those in the last section, with performances stabilising around the 83 hour mark. Again, including sample \#987 causes a much larger margin (4,634 fluorescent units at 90 hours) as the SVMs try to reduce $\xi_{987}$ to minimise the objective function, and its inclusion also increases robustness. This time, thresholds are lower (10,699 at 90 hours) due to the readings being reduced by the averaging. It could be argued that when excluding \#987, the thresholds and margins are still increasing slightly during the final few hours, and this is likely due to the second highest readings often being from replicates whose curves are still increasing. The threshold and margin at 90 hours are 10,390 and 3,931 fluorescent units respectively.

It has been argued that the inclusion of sample \#987 distorts the threshold so that it does not accurately reflect the data. This, coupled with the fact that the classification of this sample cannot be guaranteed, suggests that removing the sample from subsequent analyses might provide more accurate results. It remains to compare the remaining two sets of SVMs in \ref{singletime}(b) and \ref{singletimeavg}(b) then, but there is not a lot to choose between them.  They both result in six classification errors towards the 90\textsuperscript{th} hour, although there are two false positives and four false negatives for the highest fluorescence reading SVMs compared with one false positive and five false negatives for the averages of the two highest readings SVMs. The performance of these tests will deteriorate if final readings are taken before around 83 hours so to be confident of good results, it would probably be unwise to reduce the RT-QuIC time from a 90 hour total.

One notable difference is that for the maximum reading SVMs, the threshold appears to have plateaued by the 90\textsuperscript{th} hour, but this seems not be the case for the average of the two highest reading SVMs. It could be argued therefore that 90 hours is only sufficiently long to get optimal results for the maximum reading SVMs, whereas the averaging SVMs may need longer. However, as the robustness at these times are similar for both sets of SVMs, a clear ``winner'' cannot be identified.

Note that if the thresholds in Figures~\ref{singletime}(a) and \ref{singletimeavg}(a) are used for tests on the corresponding readings but sample \#987 is omitted, then the tests will be 100\% specific, (with the exception of readings at 74.5 hours in \ref{singletimeavg}(a)) but the SVM trained without \#987 chooses a threshold much lower and consequently false positives occur. This is due to the fact that a higher threshold would result in heavy penalties for the positive cases whose fluorescence does not increase. It may be worth considering increasing the threshold dictated by the SVM to obtain a more specific test. The thresholds are optimal in the sense that the sum of the distances of the misclassified sample readings from the threshold is minimised, but other considerations such as a preference for greater specificity cannot be recognised by SVMs and so deviating from this optimal threshold can be justified.

Sparsity induction into the training sets is meaningless here, as the training set has only one dimension and so cannot be reduced further.

\section{Tests on alternate feature spaces}

A test which improves upon the sensitivity and specificity of the null test has not been found, although it can arguably be bettered by adjusting the threshold. It is clear that the tests considered so far discard a huge proportion of the data and so the potential of using the whole data to provide a stronger test is not yet known.  A sensible progression then is to train a support vector machine on the whole of the fluorescence data, to observe if multiple readings can help distinguish further between the two classes, this is the focus of the following section.

This is first done one the entirety of the fluorescence data, without modification, but due to overfitting this approach fails. Instead, a variety of methods are explored that attempt to condense the data into just a few dimensions that summarise the nature of curve, namely parametrisation, and extracting derivative information using finite differences.

\subsection{Choosing data within samples}\label{subs:condense}
Firstly, a problem arising from the arbitrary order of the replicates must be resolved. For a given sample, it would be wrong to use all four replicates to construct one single observation of $4\times 181 = 724$ dimensions because there is no meaningful order in which to put the replicates, as they are simply four instances of an identical experiment.  Arbitrarily ordering them would created a meaningless connection between the $j$\textsuperscript{th} replicate in each sample and so should be avoided. A way must be devised to summarise the replicates within each sample.

Selecting a replicate randomly is clearly flawed as it could result in positive samples being represented by a ``flat'' replicate, despite other replicates from the same sample showing strong evidence of aggregation, and thus the major distinction between the two classes will be lost. Taking averages over the replicates does not seem sensible either. Suppose that for a positive sample, three replicates are flat and one shows a strong increase (for example \#3385), then averaging would dilute the significance of this single increasing curve because the flat readings would lower the average, bringing them closer to the negative samples. The overall effect is that observable differences between classes would diminish.

\begin{figure}[h]
\centering
\subfloat[non CJD diagnosis]{\includegraphics[width=0.5\linewidth]{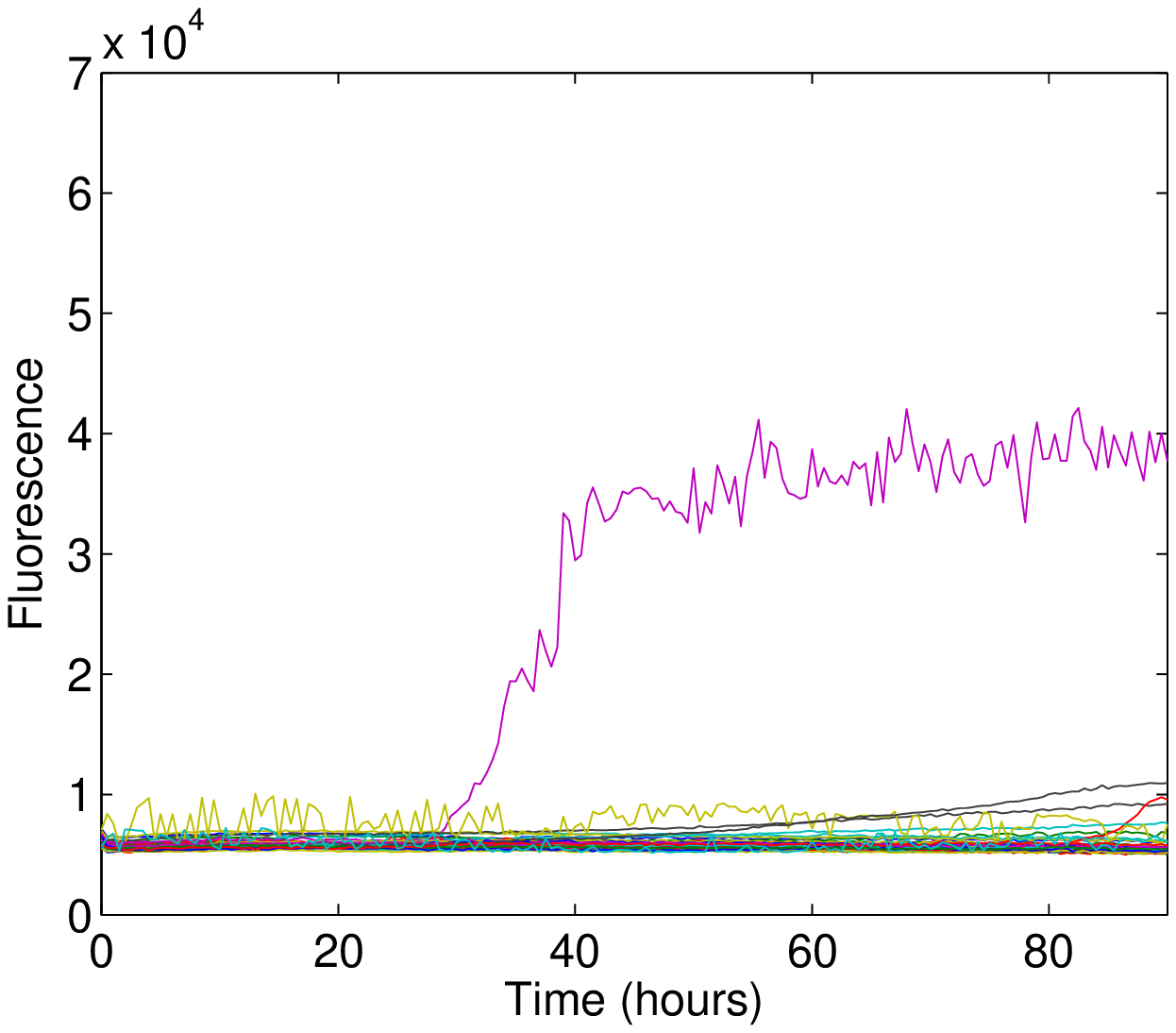}}
\subfloat[CJD diagnosis]{\includegraphics[width=0.5\linewidth]{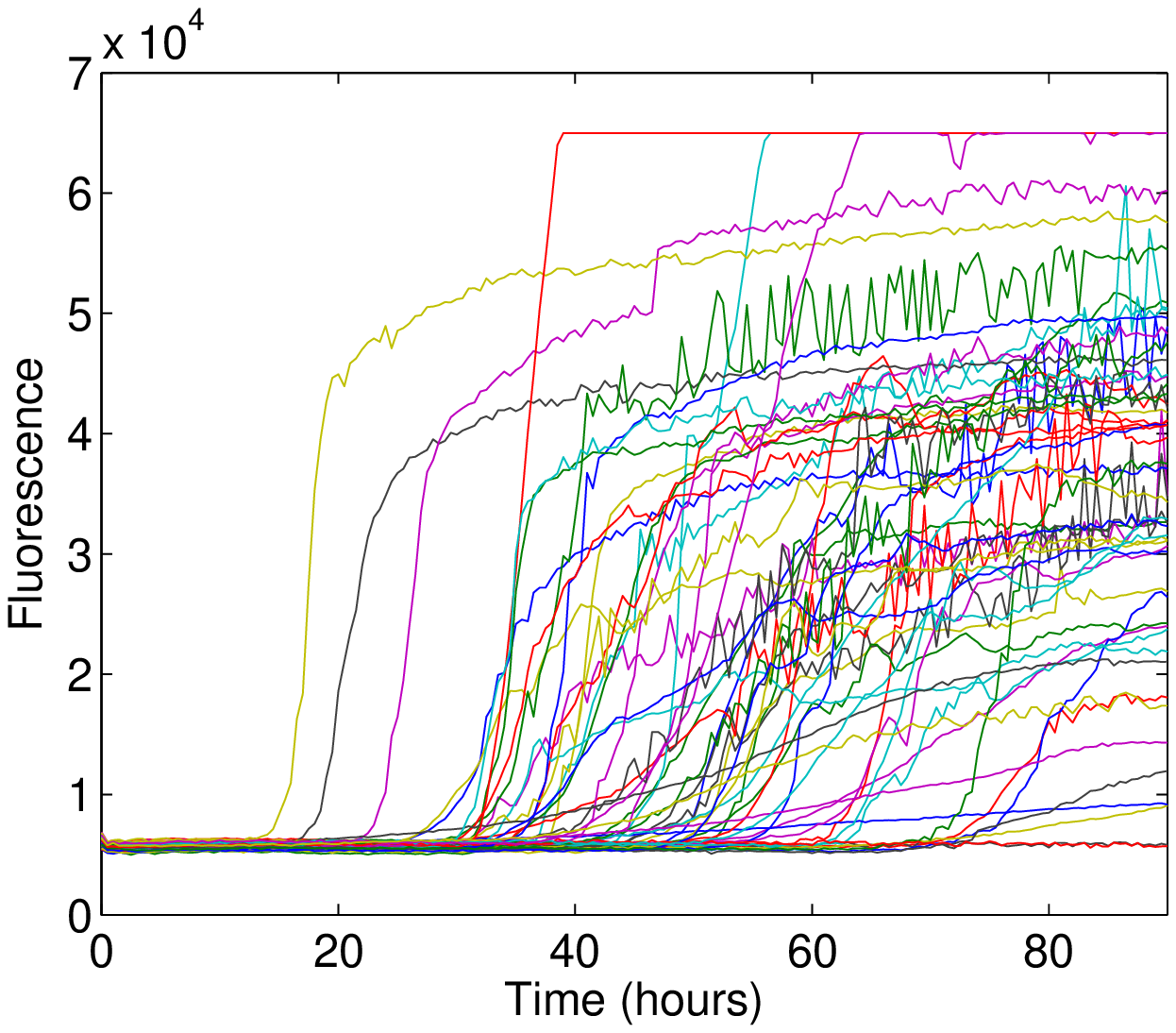}}
\caption{Curves for replicates yielding maximum total increase in fluorescence at 90 hours.}
\label{rawdataplot}
\end{figure}

A sensible heuristic is to only use data from the replicate which has the greatest fluorescence reading at 90 hours. That is, the chosen replicate for sample $i$ is
\begin{equation*}
\arg\max_{j=1,2,3,4}\{f_j^i(90)\},
\end{equation*}
where $f_j^i(t)$ is the fluorescence at time $t$ for replicate $j$. This ensures that if any significant increase in the curve occurs it is strongly represented and if no increase occurs across replicates then the maximum difference will not be significant and it will not warp the representation of the sample. For reference, this summary method shall be called the \emph{maximum total increase} heuristic. See Figure~\ref{rawdataplot} for plots of the curves selected by this heuristic for each class.

\subsection{Training on the whole length of the data}\label{robustfluoread}
Using the maximum fluorescence increase heuristic, a SVM is trained on the full length of the data. Encouragingly, the test is 100\% sensitive and 100\% specific. However, performing cross-validation on the SVM defined by the maximum total increase replicates with $C=\infty$ and $D=0$, shows that the associated test has a pseudo-sensitivity of $42/55$ and a pseudo-specificity of $49/53$, or simply $(\frac{42}{55},\frac{49}{53})_{R}$. Given that the test has a fit of $(\frac{55}{55},\frac{53}{53})_{F}$ and that the fit is an upper bound on robustness, the test is not optimally robust, and indicates that some level of overfitting is present in the test trained on the full training set.

So the test formed from training a SVM on the whole fluorescence data (with a suitable choice of replicate) has a weaker robustness---both absolutely and relative to the fit---than those tests trained only on the final reading (above), despite having 180 more dimensions and consequently this test can not be considered an improvement.

This result should in fact not be surprising as there are more dimensions than obvservations in the training set. This practically guarantees separability because there is so much space in which to seek separability, compared with the number of observations that restrict it.

A plausible remedy to this overfitting is to induce sparsity into the weight vector $w$, so that the SVM only uses features (fluorescence readings) which are necessary for separation. However, it is clear that the curves are subject to random variation, not only with regards to the noise at each measurement in time, but also the general progression of the curve over time. This becomes obvious when viewing Figures~\ref{quad1} and \ref{quad-1} given the variation present within each sample. From this, it becomes easy to convince oneself that this approach is futile, as the noise is exploitable by the SVM in its attempt to find a separating hyperplane. Analysis confirms this - separability requires a minimum of 25 dimensions, and the robustness of the feature space is poor.  However, there may still be information contained in the earlier readings which cannot be extracted by simply stacking the readings, such as the rate of change of the fluorescence levels.

The SVM analysis above proceeds by simply stacking the measurements from each observation into one long vector. Consequently, re-ordering these dimensions in the vector will have zero effect on any aspects of the analysis, because the temporal structure of these vectors is not recognised by the SVM. All dependencies of the readings on adjacent and other local readings are lost. This next section attempts to deal with this by identifying attributes of the curve that in some way measure the relationship \emph{between} the elements within each vector, rather than treating them independently.
We therefore seek low-dimensional feature spaces which, unlike sparsity induction, do not discard large portions of the data, but tries to summarise important attributes of it. The idea is to eliminate noise by retaining only the essential information, that which can be used to discriminate between the two classes.

Recall that sample \#987 has been removed, so for comparison with the null test, take $(\frac{50}{55},\frac{52}{52})_F$ as the benchmark fit.

\subsection{Parametrisation}\label{sec:params}
In this section we explore the possibility of a parametrisation of the curves. This is done by fitting some function to the curves by optimising the parameters of that function by a least-squares approach. The space of these parameters can then be used to train an SVM.
Each parameter in a given model will describe an attribute of the profile of the curve with a single value, vastly reducing the dimension of the problem, thus providing less opportunity for the SVM to exploit noise.

The most basic way to parametrise the curves would be to perform a simple linear regression on each curve, which would provide a two-parameter feature-space with which to train an SVM. However, this is unlikely to be very helpful; it is clear that most of the positively classed curves are not linear and the only important aspect of a curve it would capture is whether a significant increase has occurred. As this can be observed using the final fluorescence measurement, linear regression is unlikely to reveal anything new.

A more flexible approach is to approximate the curves piecewise-linearly. As discussed, we can roughly identify three major states in the curve profile for those which aggregate. 
By approximating these stages with three connected lines using a least-squares approach, a simplification of the curve can be achieved which has very clearly defined properties, namely the gradient and intercept for each line.

\begin{figure}[h]
\centering
{\includegraphics[width=0.6\linewidth]{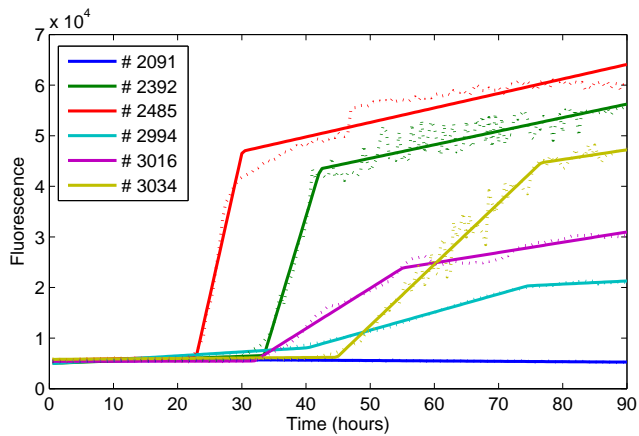}}
\caption{Piecewise linear regression with two breakpoints on a selection of samples.}
\label{linplot}
\end{figure}

The location of the two \emph{break-points}, where the lines are allowed to change gradient, can be chosen in different ways: they can be fixed, for example at 30 hours and 60 hours, but this would be highly restrictive, and it is clear from viewing the curves in Figure~\ref{rawdataplot} that this would not accurately reflect many of the curves; they can be decided heuristically, for example by using the two times where the gradient is thought to have changed the most (using second derivative approximations, see Section \ref{sec:approxs}); or most flexibly, they can be decided by absorbing their freedom using the least-squares minimisation \cite{Hastie09,Kim09}. Here the latter approach is chosen, and the resulting approximations are shown in for a selection of samples in Figure~\ref{linplot}.

It is easy to verify that there are 6 degrees of freedom in this approximation of the curves, and these can be parametrised in different ways, with an appropriate combination of intercepts, slopes and the locations of the breakpoints of the `curve'. Various such parametrisations were tried, but none improved the fit of the original test. Further, extracting the most useful parameters via sparsity induction invariably results in a single parameter being kept, and this is invariably a proxy for the final fluorescence value, such as the intercept at 90 hours, or the slope of the middle line. From this it is reasonable to conclude that additional information provided by parametrising piecewise-linearly does not help find a test to rival the single dimension tests.

Attempts were also made to fit various non-linear functions to the curves via least squares regression, with the intention of using the space of the parameters of these functions to train SVMs.

\begin{figure}[h]
\centering
{\includegraphics[width=0.6\linewidth]{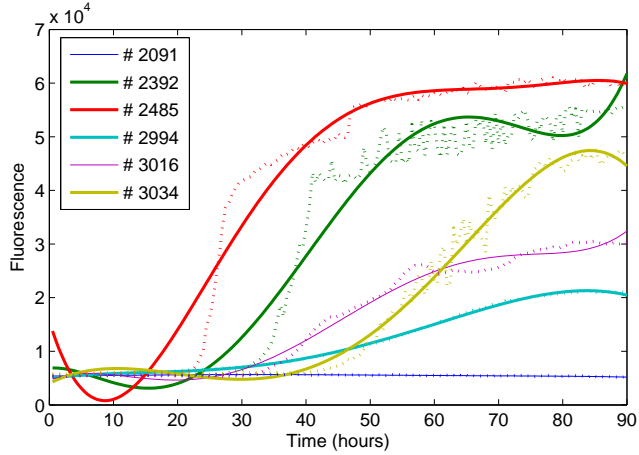}}
\caption{Polynomial regression of order 4 on a selection of samples.}
\label{polyplot}
\end{figure}

Polynomial approximations were tried, but these were a poor fit for lower orders, and convergence of polynomials of higher orders was frequently unstable, so the polynomial approach was abandoned. Figure~\ref{polyplot} displays the fits for polynomials of order 4, and it is clear that for sharply rising curves, the regression fails to capture the trend.

Another family of non-linear functions, called Sigmoid functions, were tried. These are used to describe a monotonic curve with upper and lower asymptotes and intermediate growth - most of the positive cases match this description and it seems plausible that the flat curves which do not follow this trend could be modelled by degenerate or distorted cases of the sigmoid function (linearly, for example, with the lower and upper asymptotes equal). Unfortunately, it was not possible to find a sigmoid function with enough flexibility to accommodate the many idiosyncrasies of each curve adequately. Given that the support vector approach requires the space in which the observations are expressed to be identical for \emph{all} observations, using different functions (and therefore a different parameter-space) is not feasible, and so this approach was also abandoned.

\subsection{Local approximations to the curve using finite differences}\label{sec:approxs}
One aspect of the data which cannot be easily interpreted by stacking the dimensions is derivative information along the curve, in particular the gradient and the curvature of the curves at different times. If this information is obtained locally at regular intervals it can form a training set to be used for SVM analysis. However, this approach would also involve stacking the dimensions and so the problems associated with stacking, such as wrongly assuming independence of adjacent measurements, will still be present. Instead, by extracting certain potentially significant values---namely, the maximum and minimum fluorescent values, the maximum and minimum first- and second-derivative estimates, and also the times that these extrema occur---a feature space of just 12 dimensions can be formed (potentially reduced via sparsity induction) which may be of use for SVM training.

As no functional representation for each curve could be found, gradient information will be approximated by other means---using \emph{finite differences}---the differences between two points on a curve a distance $h$ apart \cite{Hildebrand68}. There are a variety of finite difference methods available. This paper uses the \emph{central} difference, defined by
\begin{equation}
\delta_h f(t)=f(t+\tfrac{1}{2}h)-f(t-\tfrac{1}{2}h),
\end{equation}
where $f(t)$ is the fluorescence at time $t$. The central difference is preferred as it does not rely solely on measurements that follow it, so for large $h$, the approximation is less local, and more stable. The gradient, or rate of change, approximation may be calculated by dividing the difference by distance $h$, so we have
\begin{equation}
f^\prime(t)\approx\frac{\delta_h f(t)}{h}=\frac{f(t+\frac{1}{2}h)-f(t-\frac{1}{2}h)}{h}.
\end{equation}
Second derivatives, which measure curvature, can be approximated by
\begin{equation}
f^{\prime\prime}(t)\approx\frac{\delta_h^2 f(t)}{h^2} = \frac{f(t+\frac{1}{2}h)-2f(t)+f(t-\frac{1}{2}h)}{h^2}.
\end{equation}
With respect to the fluorescent data, first derivative approximations are measured using \emph{rfu per hour} and second derivative approximations are measured using \emph{rfu per hour-squared}. Measurements are taken every half-hour, so $h$ must be a multiple of a 1-hour interval in order that there are measurements available for calculations.

The maximum (minimum) first derivative approximation should correspond to the time on the curve where the steepest positive (negative) gradient occurs. The maximum (minimum) second derivative approximation should correspond to the time on the curve where the sharpest upturn (downturn) occurs.

Unfortunately, finite differences are very sensitive to noise and this can severely distort the approximations. This is because each approximation is calculated using just two observations, so significant changes between adjacent observations due to noise can cause marked differences between adjacent approximations, even if the trend suggests they should be equal or close. This issue can be partly resolved if $h$ is large enough, as the noise between $t-\frac{1}{2}h$ and $t+\frac{1}{2}h$ is simply bypassed. However, this will work only where the change dictated by the trend is greater than the noise, which is certainly not the case wherever a curve appears to be flat. Also, if $h$ is too large then the approximation cannot be guaranteed to be local enough for sufficient accuracy.

For the fluorescence data, preliminary analysis shows that only using large $h$ is not enough to bypass noise and gain sufficient accuracy to capture the trend. An additional method to deal with noise, which can be employed in conjunction with different $h$ sizes, is to \emph{smooth} the curves, a process which attempts to identify the underlying temporal trend of longitudinal data series \cite{Chatfield04}. The smoothed data should approximate the progression of this trend, as if no noise is present. A popular and easy to implement smoothing method is the \emph{moving average filter}, which averages a succession of contiguous, fixed-size subsets of longitudinal data, to provide an estimate for the central point in the subset. The set-size must therefore be odd, so that a central point exists. If the set size is $k$, then the \emph{smoothed} estimate for observation $f(t)$ is
\begin{equation}
g(t)= \frac{1}{k}\sum_{r=\frac{k+1}{2}}^{m-\frac{k-1}{2}} f(r),
\end{equation}
where $m$ is the length of the series. For the first and last $\frac{k-1}{2}$ points where there is not enough data to average over $k$ values, the size of $k$ is reduced accordingly. The choice of $k$ is important; too small and the curve will not be sufficiently smoothed, yet too large and it will not adequately capture the trend, so a compromise must be made.  The ``best'' $k$ is subjective, and can be gauged by visually comparing the smoothed curve against the raw.

\begin{figure}[ht!]
\centering
\subfloat{\includegraphics[width=0.75\linewidth]{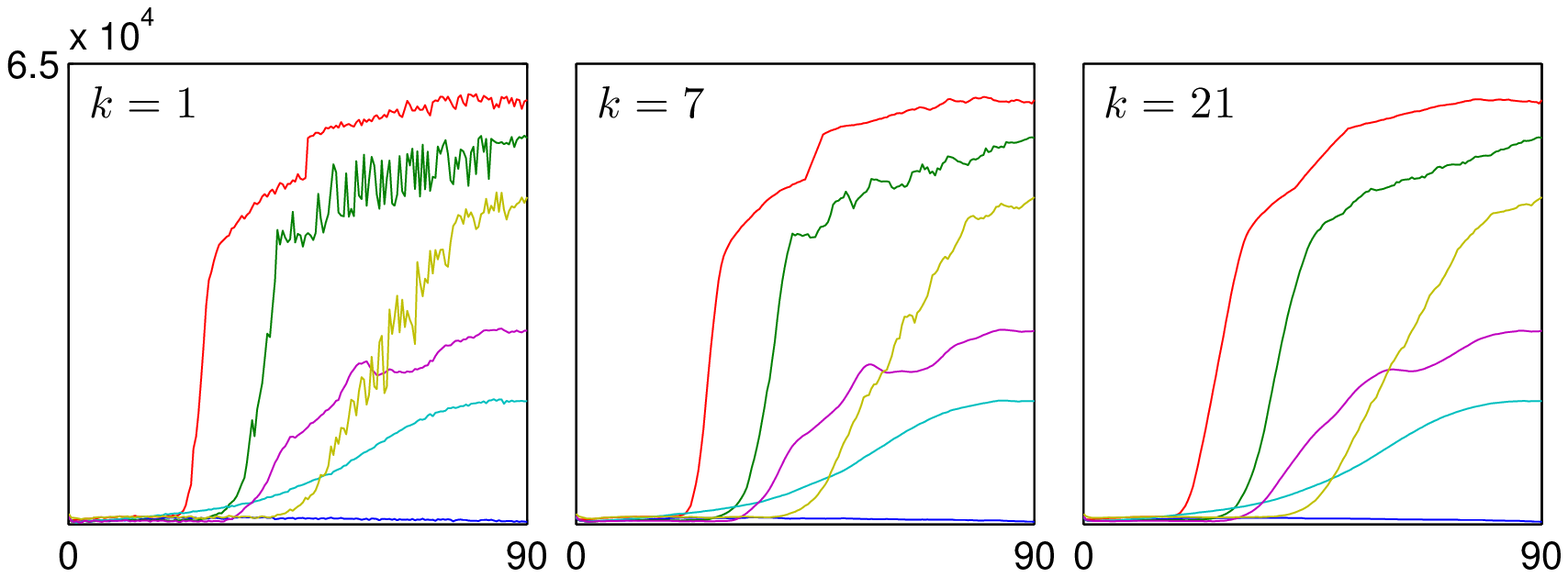}}\\
\vspace{-0.5cm}
\subfloat{\includegraphics[width=0.75\linewidth]{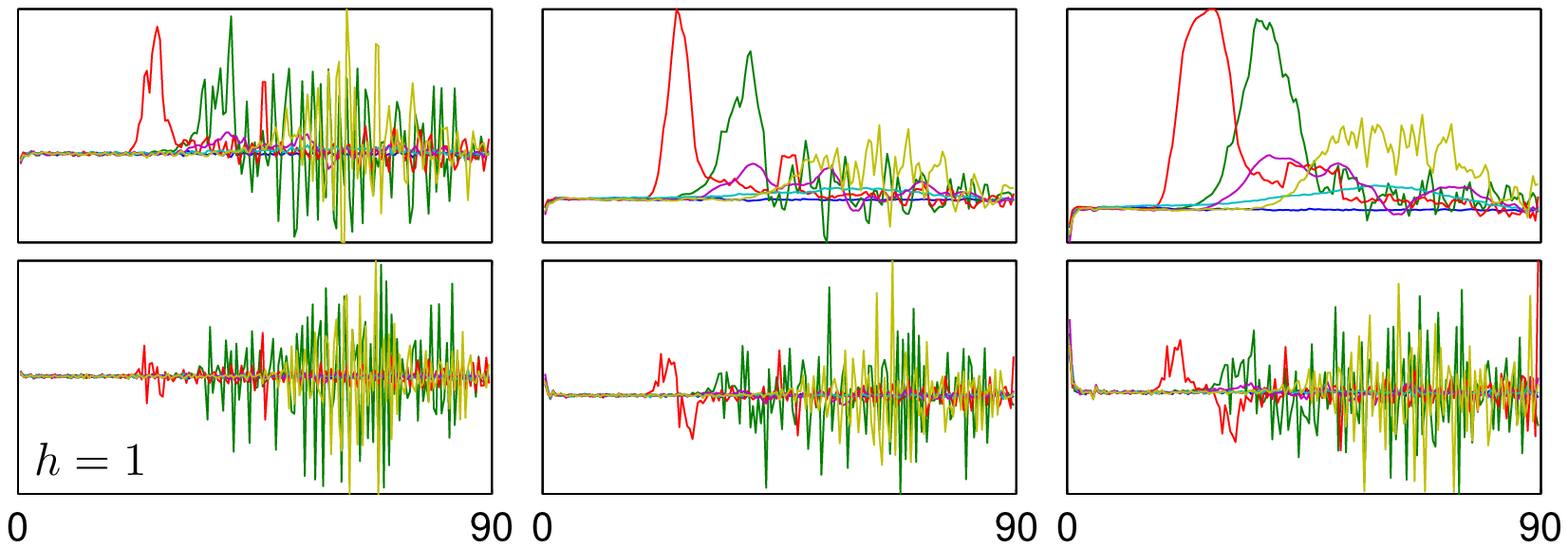}}\\
\vspace{-0.5cm}
\subfloat{\includegraphics[width=0.75\linewidth]{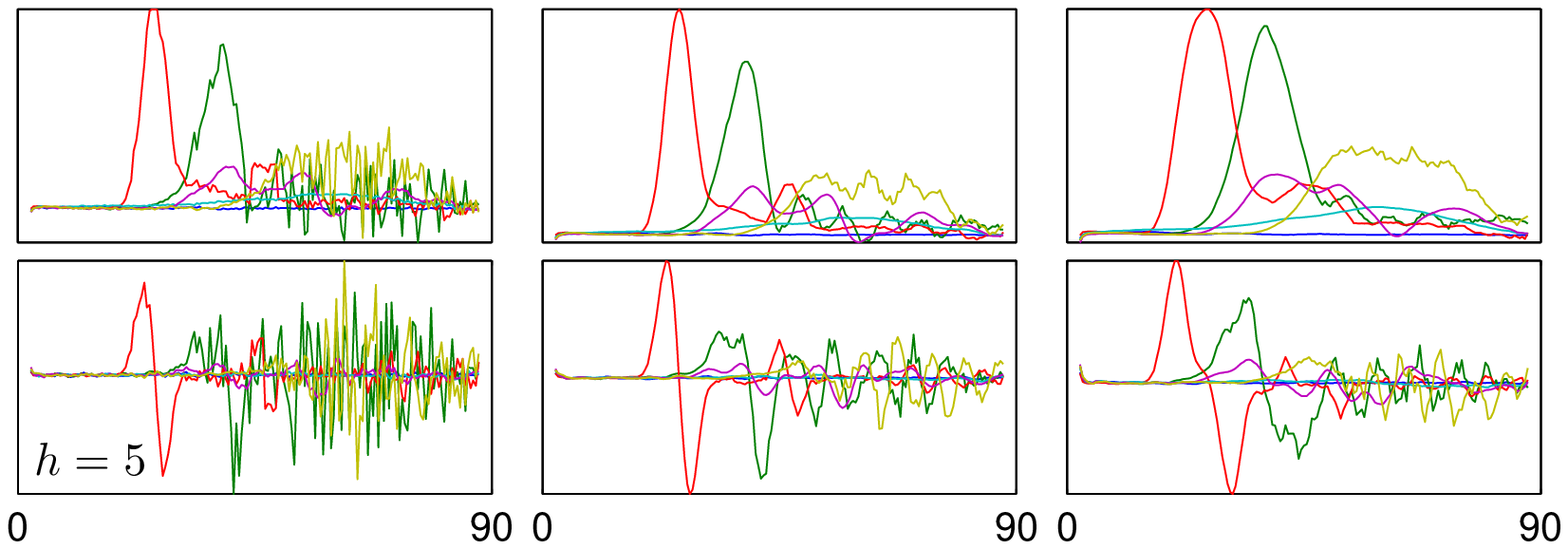}}\\
\vspace{-0.5cm}
\subfloat{\includegraphics[width=0.75\linewidth]{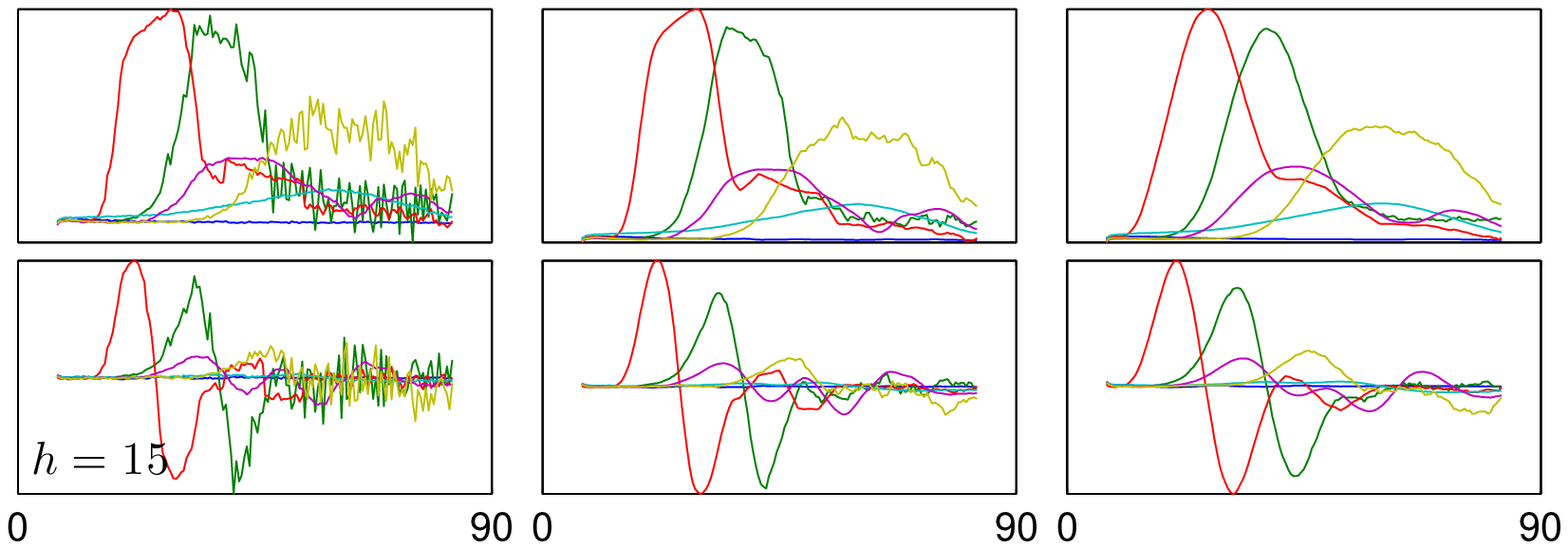}}\\
\caption{Smoothed fluorescence readings for various values of $k$ and $h$ and a selection of samples, from 0 to 90 hours. The panels on the top row show the effects of smoothing for different values of $k$. Note that $k=1$ corresponds to no smoothing. The remaining 18 panels show the first and second derivatives approximations of the smoothed data from the corresponding top panel, for different $h$. Upper sub-panels are the first derivatives and lower sub-panels are the second derivatives. It should be noted that the scale of the vertical axes vary for each panel, but as this information is irrelevant for observing the impact of smoothing and finite differencing, the scales are omitted to avoid clutter. }
\label{smoothed}
\end{figure}

The smoothed data for a selection of samples (chosen as representative examples to avoid visual clutter), are plotted in Figure~\ref{smoothed} for different values of $k$. Also plotted are the first- and second-derivative approximations of this smoothed data via finite differences, $\frac{\delta_hg(t)}{h}$ and $\frac{\delta^2_h g(t)}{h^2}$, for various values of $h$.

A brief analysis of Figure~\ref{smoothed} highlights the value in smoothing and justifies its use. Firstly, the derivative information gained from $k=1$ and $h=1$ does not reflect the trend well because, for example, after the steep rise for the green sample the trend suggests a steady linear increase which would dictate that the first derivative is positive constant and the second derivative is zero, but instead the first- and second-derivatives oscillate wildly about zero. As $k$ and $h$ increase these oscillations are greatly reduced, and the locations of the extrema become more consistent with the trend, so that, for example, the maximum second-derivative is where the curve starts to rise, rather than afterwards where the increase appears to have stopped. However, the curves tend to be ``flatter'' for large $k$, so the first and second derivative approximations may be closer to zero than the trend would suggest.

This assumes, of course, that the non-smooth nature of the curves is in fact due to noise, and not perhaps some oscillatory behaviour which has some relevant meaning. This possibility is not examined in this paper.

As expected, training SVMs on the vectors created by stacking the first- or second-derivative approximations results in overfitting. Table~\ref{stacksmoothfindiff} presents the robustness measures for a selection of such SVMs and shows that the robustness is consistently poor.

\begin{table}[h]
\centering
\begin{tabular}{l|ccc|ccc}
 & \multicolumn{3}{c}{first-derivative} & \multicolumn{3}{c}{second-derivative}\\
\hline
 			& $k=1$ 			& $k=7$			& $k=21$ &  					 $k=1$ 			& $k=7$			& $k=21$\\
\hline
 $h={\phantom 0}1 $ & $(46,48)$ 		& $(41,47)$ 	& $(42,47)$		& $(25,45)$ 	& $(26,45)$ 	& $(34,46)$\\
 $h={\phantom 0}5$ & $(42,48)$ 		& $(39,46)$		& $(40,48)$	 	& $(26,50)$ 	& $(30,46)$ 	& $(38,46)$\\	
$h=15$ 				& $(44,51)$ 		& $(42,47)$		& $(40,48)$		& $(40,48)$ 	& $(38,46)$ 	& $(44,48)$
\end{tabular}
\caption{Robustness of SVMs trained on first- and second-derivative approximations with smoothing parameter $k$ and finite difference parameter $h$, interpreted as (\emph{number of true pseudo-positives, number of true pseudo-negatives}). The denominators (55 total positive cases and 52 total negative cases) are omitted for clarity.}
\label{stacksmoothfindiff}
\end{table}

Instead, a summarising set of features is used, and this is taken to be the 12-dimensional feature space created by identifying the extrema of the fluorescence values and the first and second derivatives, together with their times. The set is tested for its use in SVM classification, and this can be done for differing values of $k$ and $h$ to assess if any particular combination works better than others. The appropriate optimisation problem for $C=\infty$ and $D=0$ is infeasible (implying non-separability) for many, but not all, $k$ and $h$ combinations. The very fact that separability can occur in just 12 dimensions, rather than the minimum of 25 dimensions needed for separability in the stacked training set, indicates that summarising the curves in this way allows for a more meaningful interpretation of the data by the SVM.

\begin{figure}[ht!]
\centering
\subfloat[\# false positives]{\includegraphics[width=0.95\linewidth, height=5cm]{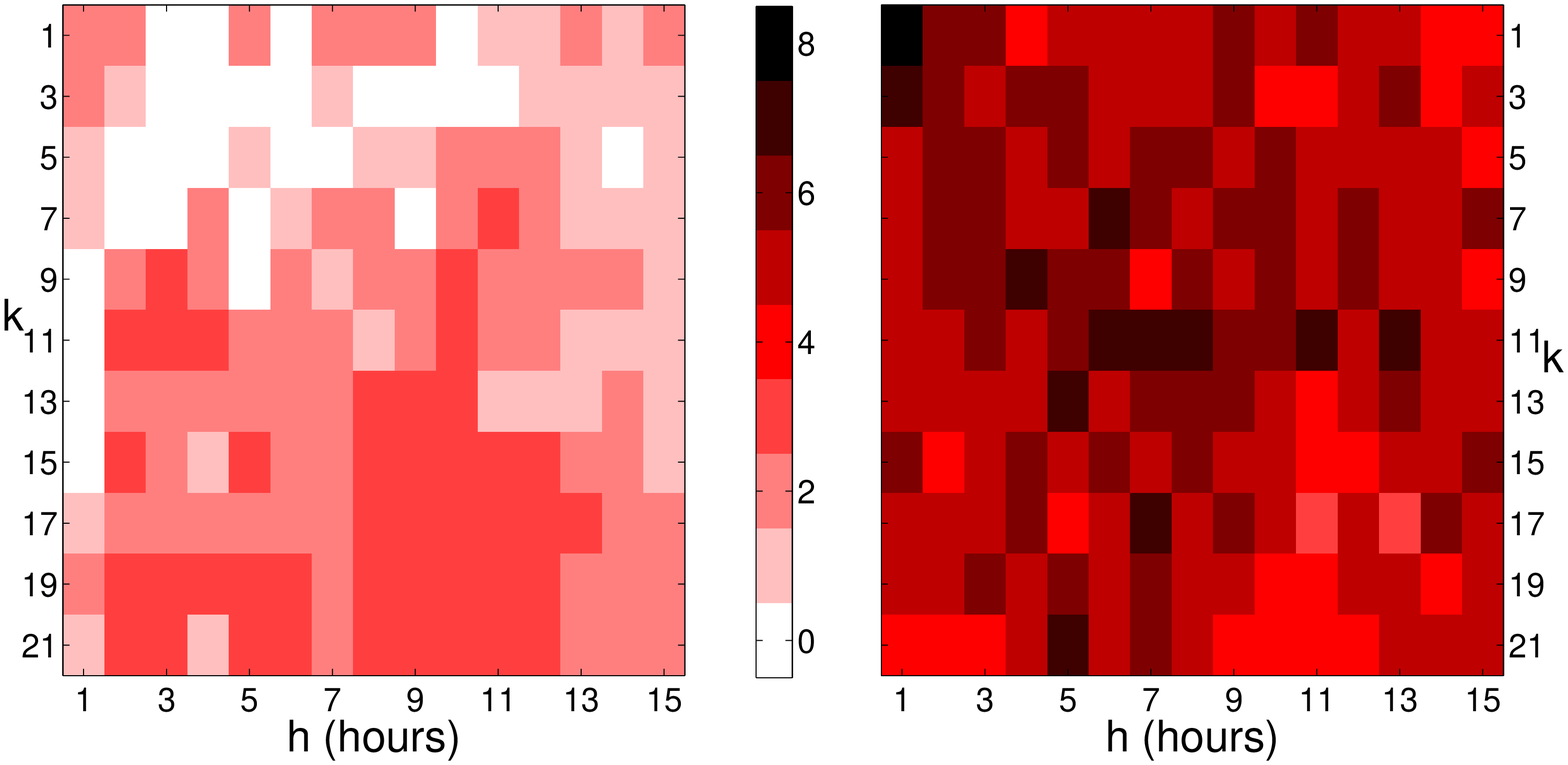}}\\
\subfloat[\# false negatives]{\includegraphics[width=0.95\linewidth, height=5cm]{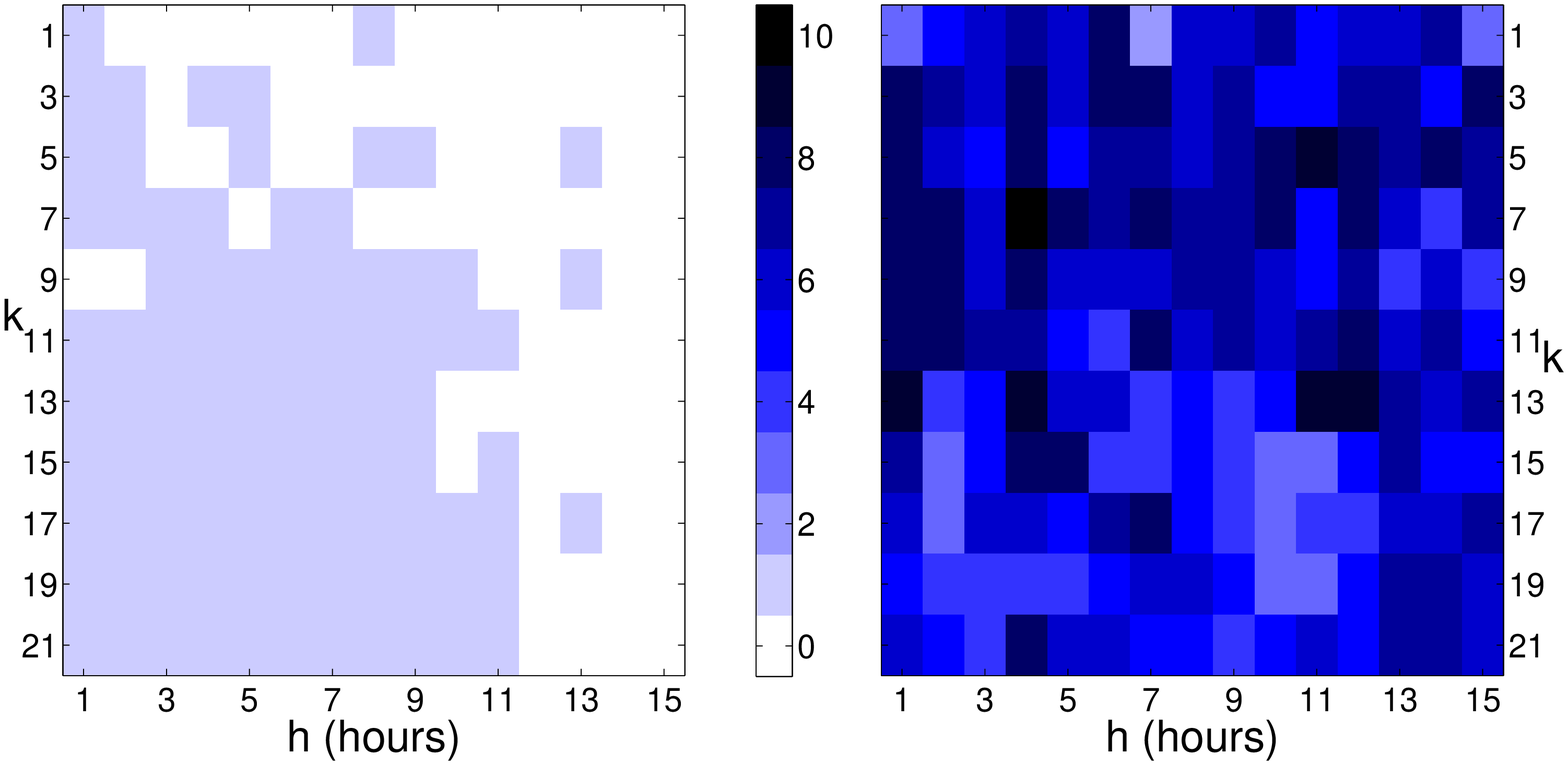}}\\
\subfloat[\# total errors]{\includegraphics[width=0.95\linewidth, height=5cm]{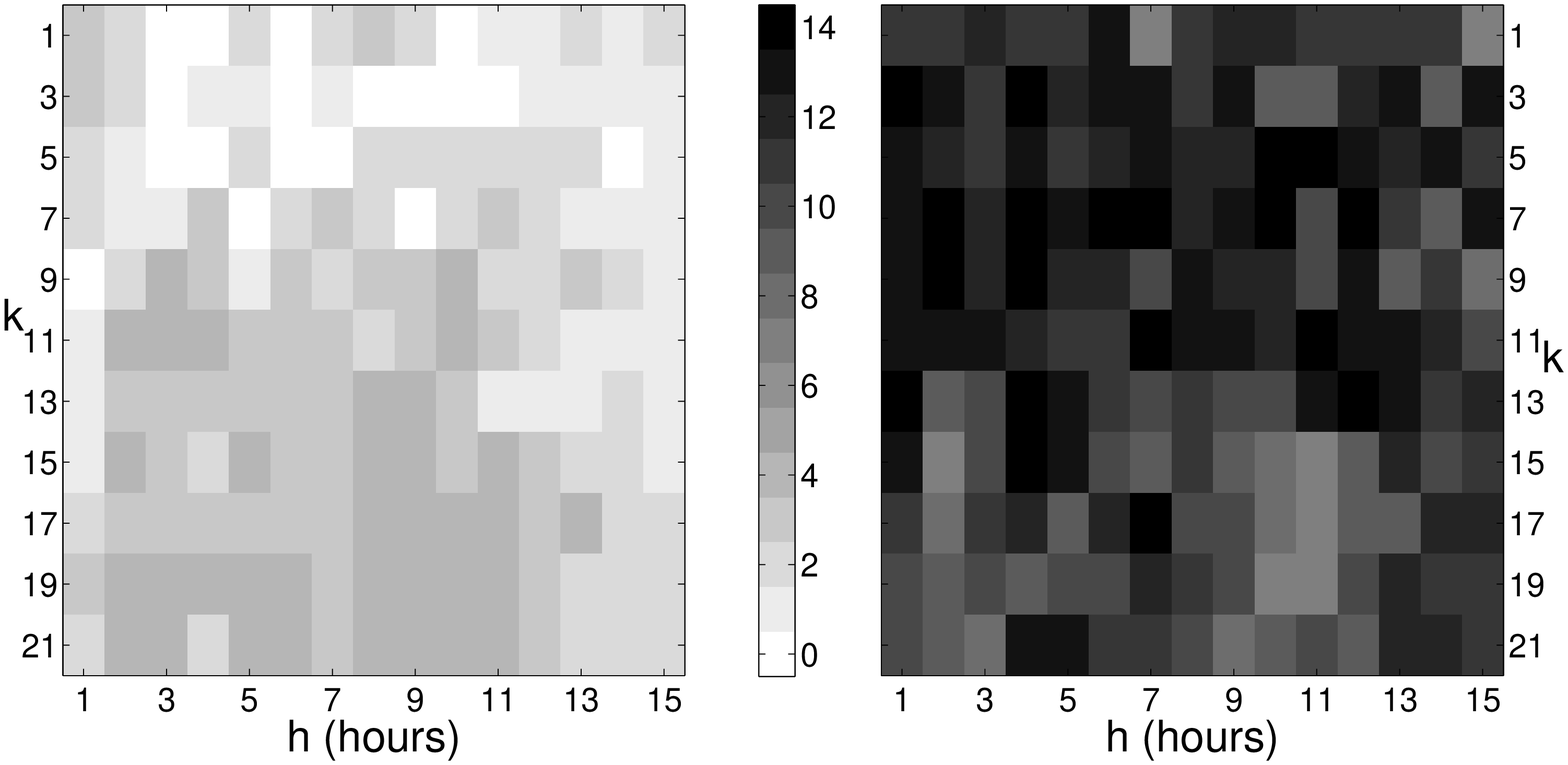}}\\
\caption{Colour plots indicating number of classification errors of SVMs trained on a set comprising of the extrema (over all hours) of the fluorescence values and the first and second derivatives, together with their times. $15\times 11=165$ such training sets were created (recall that $k$ must be odd), each the result of a unique combination of $k$ and $h$, the moving average parameter and finite difference parameter respectively. Lighter shades indicate better performance. Plots on the left are actual classification errors and plots on the right are pseudo-classification errors via cross-validation.}
\label{smoothedcolour}
\end{figure}

For the remaining non-separable combinations, an appropriately sized $0<C<\infty$ is used ($C\approx 10$ is enough large enough to ensure that only the vectors preventing separability have $\xi_i>0$). The classification performance of these SVMs for varying $k$ and $h$ are presented in Figure~\ref{smoothedcolour}, where the number of misclassifications determines the shade of the colour.

Analysing these graphs, a slight tendency for low $k$ to yield less classification errors can be seen, and for high $k$ to yield less pseudo-classification errors, and consequently, where there are very few classification errors (light shades), the robustness tends to be weaker, indicating that near-separability is achieved only by overfitting to the data. This suggests that there is not an optimal pairing for $k$ and $h$, because improving the fit reduces the robustness.

One potential explanation for this is that the efforts to reduce and bypass noise are not particularly effective, but this goes against the apparent value of these methods that is demonstrated under a graphical interpretation from Figure~\ref{smoothed}. Another explanation is that the smoothing does improve the accuracy of the first- and second-derivative approximations, but the SVMs are not using these for separations. This latter explanation can be investigated by inducing sparsity into the weight vector and observing which features the SVM selects. For this feature space, sparsity induction is actually more valid than in the stacked case as the SVMs will no longer be dismissing single frames in time, but instead more general aspects of the whole curve which may be irrelevant to the differences between the classes.

The results are enlightening: when $D$ is large enough to perturb the solution, the SVM immediately rejects the dimensions corresponding to the times of the extrema, suggesting that these dimensions have no real value for discriminating between cases. The number of classification errors remains roughly the same, although the balance shifts towards one or two more false positives and one or two less false negatives. The robustness in general is improved, which again points toward overfitting for $D=0$.

If $D$ is completely dominant, the SVM always selects the feature corresponding to the maximum fluorescent level, and it always has a fit of $(\frac{50}{55},\frac{52}{52})$ and a robustness of $(\frac{50}{55},\frac{52}{52})$, regardless of the choice of $k$ and $h$. In fact, as $h$ is a parameter of the finite differences, it has no effect at all on the maximum fluorescence dimension, so $h$ need not be considered when examining only this dimension.

If a SVM is trained on either the maximum first derivative feature, the maximum second derivative feature, or the minimum second derivative feature, then the performances are quite good, but inferior to using the maximum fluorescence. This should be unsurprising, as these measures are all be correlated with maximum fluorescence; if the maximum fluorescence of a curve is high, then the gradient must increase at some point. Similarly, this increase in slope must cause the second derivative to increase. As the curves level, then the second derivative goes negative, and the minimum second derivative will correspond to this. Unfortunately, the SVM does not detect anything extra in these features or the times that they occur that eradicates false positives and false negatives and improves robustness. Consequently, they can only be considered proxies for the maximum fluorescence and so their inclusion is not warranted.

Choosing the maximum fluorescence as the single feature results in the maximum achievable robustness via finite differences and smoothing and consequently, using information from finite differences does not result in a test that supersedes the null test.

\section{Effects of additional data}\label{sec:classadditional}
All the previous analyses used only fluorescence data obtained from RT-QuIC analysis on the CSF samples from each patient. In this section, the possibility of using the additional data (sex, age, date of LP procedure) to strengthen the test is considered. Disease duration and time to LP procedure is not considered here, as this data is unavailable for non-sCJD cases.

\begin{figure}[ht!]
\centering
{\includegraphics[width=0.95\linewidth]{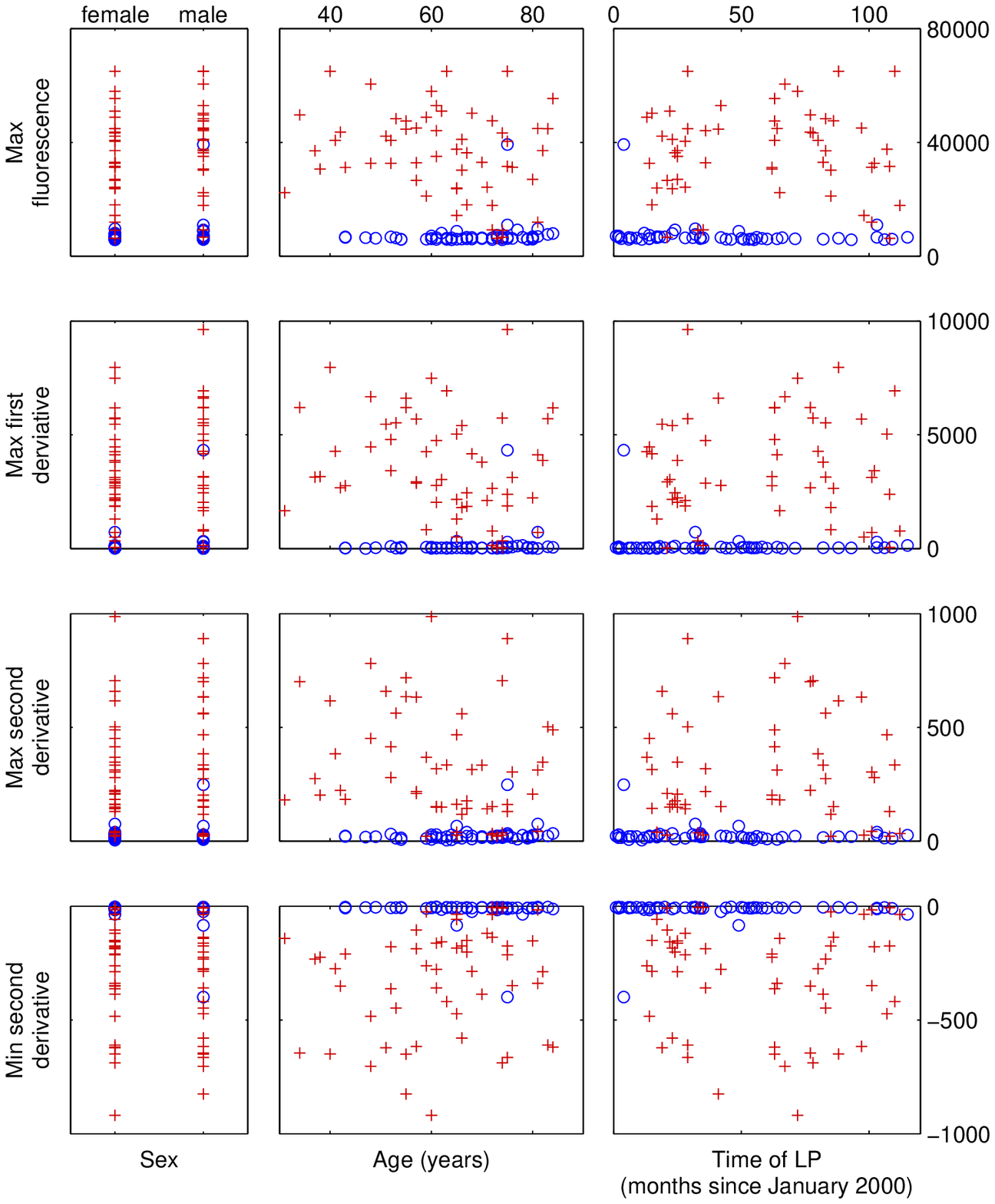}}
\caption{Scatter diagrams that plot various summaries of the curve against potential predictors and sCJD diagnosis.  Red crosses are sCJD positive cases and blue circles are sCJD negative cases. Derivatives were obtained using a smoothing factor $k=11$ and a finite difference factor $h=11$. There appears to be no correlations, and other graphs for different $k$ and $h$ show similar results.}
\label{classpatterns}
\end{figure}

This can be investigated preliminarily by plotting summary measures of the curve from Section~\ref{sec:approxs} against these additional \emph{predictors}, and observing if any correlations or patterns present themselves and these plots are displayed in Figure~\ref{classpatterns}. Sample \#987 has been reintroduced to the data for plotting.

Firstly, the plots clearly shows that age and sex alone are completely unreliable as predictors of sCJD. There is no clear segregation of these factors by sCJD subtype at all, and this conclusion can be drawn by simply ignoring the dimensions on the vertical axes and observing the distribution of positive and negative cases along the horizontal axes. Unsurprisingly, this is also true for the time to LP procedure. Consequently, if these additional predictors have no impact on the profile of the curves (vertical axes), then they will have no value as additional features in an SVM.

It is clear that no correlations are apparent between the curve profile measures and any of the predictors. A clear difference exists between positive and negative cases, and this is true for all four curve profile measures. By ignoring the data on the horizontal axes, it can be seen that the maximum fluorescence is the best indicator of these differences, proven by the fact that this feature is selected via sparsity induction in Section \ref{sec:approxs}. The remaining measures plotted here are merely proxies for the maximum fluorescence, and it is easy to see that there is a greater overlap of cases by these measures.

One small characteristic of the data to note is that there is a more distinct difference between the two cases when the patients are younger. This can most clearly be seen in the panel corresponding to maximum rfu and age, where all of the positive sCJD cases that are misclassified by the null test (those with low maximum rfu values) are at least 70 years old, and hence significantly older than the positive population average (62.1 years). There are also many CSF samples from older sCJD patients with high maximum rfu values, so this characteristic is not indicative of a correlation with age, but an increase in the variance of maximum rfu with age. As this does not constitute a separation of the data, SVM cannot exploit this characteristic to improve the reliability of the test. This would be exploitable by probabilistic classifiers which can report the probability of a case being positive---in this case the probability of an older patient with no suggestion of an increase in fluorescence would be higher than for a younger patient with a similar curve---but such classifiers are not considered in this paper.

Similar graphs plotting the times of the extrema of rfu values and of the first- and second-derivatives against the predictors show no significant patterns either. SVMs were trained with these predictors and curve summarisers, but no tests were found that improved separability without reducing the robustness. The details of these SVMs is omitted.

\section{Seeking differences between CJD types}\label{chap:type}
The null test makes use of the differences between sCJD positive and sCJD negative cases that present themselves in the RT-QuIC data. Further clinical distinctions can be made between positive cases, but whether or not these distinctions are manifest in the RT-QuIC data has not been explored. For the purposes of developing a more detailed diagnosis, this issue is addressed in the following section, using SVM analysis.

There are 55 positive cases, of which 30 are type CJD-MM, 17 are type CJD-MV and 8 are type CJD-VV. Consequently, a brief discussion on extending the binary classification techniques described in Section \ref{chap:SVMs} to $p>2$ classes is needed so that the SVM approach can be applied to this 3-class problem.

Two common approaches to multi-class classification using SVMs are \emph{one-versus-all} and \emph{one-versus-one}; see \cite{Hsu02}.  One-vs-all constructs a separate SVM for each of the $p$ classes, where the training set for each SVM is taken as one of the classes versus the remaining classes. Classification of a new observation is achieved by assigning it to the class whose SVM gives it the greatest positive distance from the hyperplane. This is the approach used here, but it soon becomes apparent that neither approach would improve the reliability of a test because no differences between types can be detected in the curve.

\subsection{Analysis on fluorescence data}

\textbf{Stacked data.}\label{stackedmulti}
The three SVMs constructed using the maximum yield training set are all linearly separable, which is highly unsurprising given the freedom in the training sets to find separability. Cross-validation confirms that this is purely noise-exploitation (results omitted), and so analysis of this stacked training set ends here.  The next step is to try derivative information obtained from finite differences.

\textbf{Analysis on curve profile measures.}\label{curveprof}
Constructing the 12 dimensional finite-differences training set in the usual way (as described in Section \ref{sec:approxs}) for a given $k$ and $h$, we can observe if there are any discernible differences between the types present in features which summarise the profile of the curve. The results show that no meaningful differences can be found. Separability is not possible in any case, the robustness measures are very often below 50\%, and the selection of features via sparsity induction is highly dependent on noise.  There can be absolutely no expectation that derivative information in the curve can be reliably used to diagnose the specific sCJD subtype.

\subsection{Effects of additional patient data}
As in Section~\ref{sec:classadditional}, this section deals with the possibility of using additional patient data to help discriminate cases. Together with the sex, age and the date of the LP procedure, further data detailing the duration of disease and the time of the LP since first symptoms is available. An additional predictor may be constructed by dividing the LP time by the duration of disease to observe directly how far into the disease the patient was when the sample was taken, as a proportion of the total disease duration.

Rather predictably, the LP time alone does not influence the sCJD subtype, but it is conceivable that LP time and the other \emph{predictors} can be used to partially explain the profile of the curve. It is worth noting that for a pre-mortem diagnosis, disease duration cannot be used as a predictor as it will not be known. As such, it may be useful to incorporate these predictors into a training set so as to augment the space of features in which to seek differences between the three sCJD subtypes, even though the profile of the curve alone does not provide enough information to be used as a diagnostic tool.

\begin{figure}[ht!]
\centering
{\includegraphics[width=0.95\linewidth]{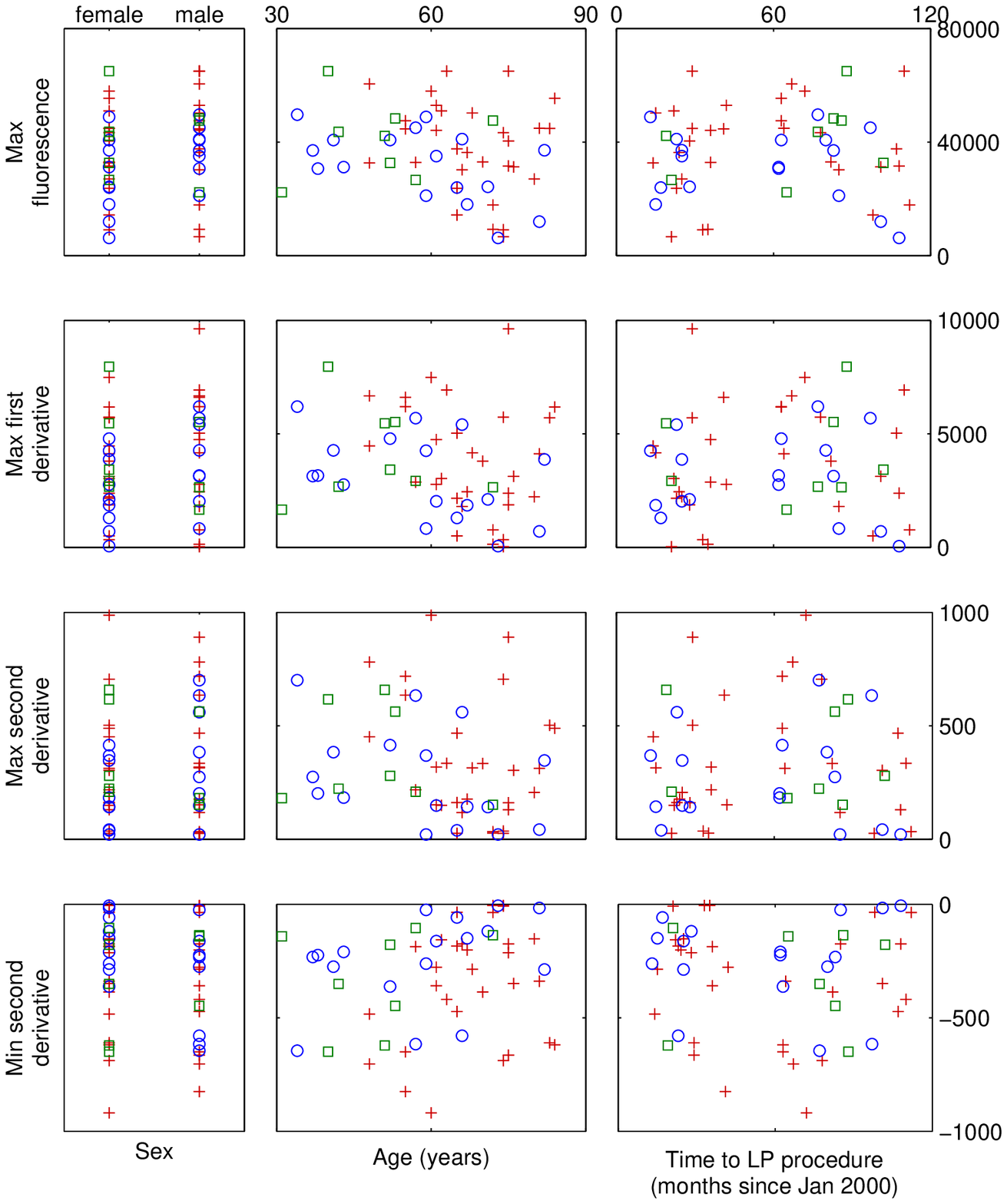}}
\caption{Scatter diagrams that plot various derivative measurements against potential predictors and sCJD type. Derivatives were obtained using a smoothing factor $k=11$ and a finite difference factor $h=11$. Red crosses are CJD-MM cases, blue circles are CJD-MV cases and green squares are CJD-VV cases. There appears to be no relationships between the measurements and any of the factors.}
\label{typepatterns}
\end{figure}

Via Section~\ref{sec:approxs}, derivative information for the curves is obtained for various $k$ and $h$, and are plotted against sex, age and the date of LP procedure to examine if these predictors do influence the curve profile. These are presented in Figure~\ref{typepatterns} for $k=11$ and $h=11$ and it is clear that no correlations or patterns are present. This is also true for all combinations of $k$ and $h$.

\begin{figure}[ht!]
\centering
{\includegraphics[width=0.95\linewidth]{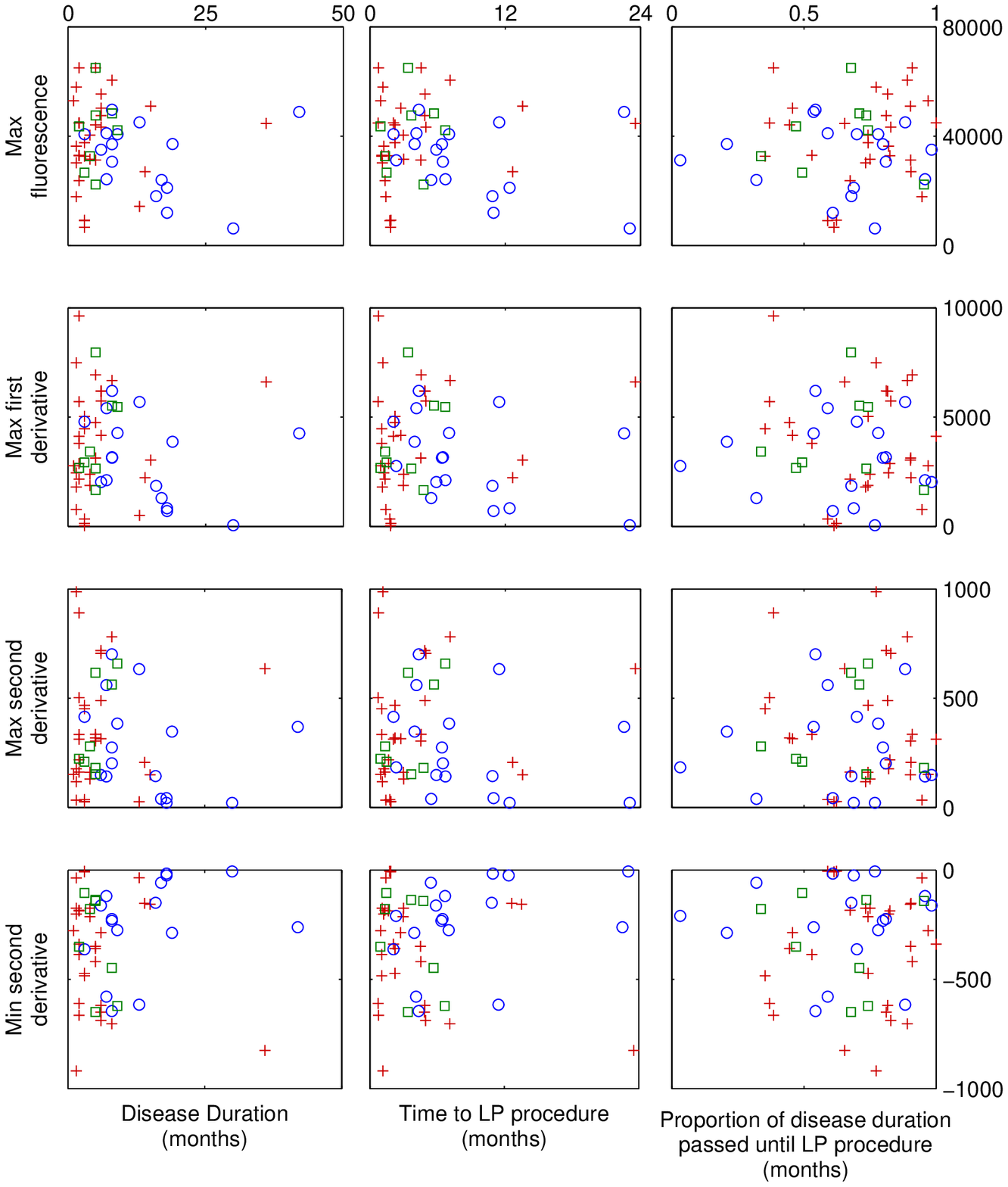}}
\caption{Scatter diagrams that plot various derivative measurements against potential predictors and sCJD type. Derivatives were obtained using a smoothing parameter $k=11$ and a finite difference parameter $h=11$. Red crosses are CJD-MM cases, blue circles are CJD-MV cases and green squares are CJD-VV cases. There appears to be no relationships between the measurements and any of the factors.}
\label{typepatternsdur}
\end{figure}

The patient data available only for positive cases is also plotted in a similar way in Figure~\ref{typepatternsdur} and again, no clear correlations or patterns emerge, which holds for all $k$ and $h$.

Of course, this pair-wise analysis of different measures is not enough to confirm that these additional predictors do not aid in detecting differences between types, but SVM analysis does confirm this. Separability is not possible and sensitivity and specificity measures in the soft margin case are poor. The overall conclusion then is that sCJD subtypes are not identifiable from the data.

\clearpage

\section{Conclusions}

\subsection{Distinguishing between sCJD and non-sCJD cases}
The most reliable tests for sCJD can be formed simply using the rfu readings at 90 hours.  Training SVMs on any additional data, despite increasing the fit, does not improve robustness and is therefore less reliable. However, with more data, better cross validation techniques can be used; and it is conceivable  that a different test might have both a better fit and better robustness.

Measures which summarise the profile of the curve, such as the maximum first derivative approximation, are strongly correlated with the readings at 90 hours and so provide similar, but inferior, measures of the total rfu increase. Using these measures in addition to the maximum rfu increase would not improve performance. There are no useful patterns to exploit in additional patient data such as age and sex.

The rfu readings at 90 hours can be used in different ways, but the most effective of these is not obvious. The threshold obtained when training all 108 samples on the highest rfu readings at 90 hours has the same performance as the null test, but it was shown that the dubiously classified sample \#987 was distorting this threshold and as such might not reflect the data well.  Removing this sample reduced the threshold considerably, and changed the balance of classification errors.

The null test is not significantly improved upon, but one may wish to consider using only the maximum rfu reading at 90 hours for its simplicity. In this case, the threshold suggested by SVM classification (without \#987) is roughly 10,000 rfu but this gives a less specific (but more sensitive) test. If desired, the performance of the null test can be recovered by shifting the threshold to around 14,000 rfu.

The major finding of this paper is that using the 90\textsuperscript{th} hour rfu readings is the most robust way to discriminate between classes---the calibration of the threshold is secondary. Also, it is likely that the result of a test to a new sample whose maximum reading at 90 hours is around 10,000 rfu will be considered inconclusive, as there is a degree of overlap of the classes around this level. The classification of such a sample may be postponed until further tests are carried out, and so the exact value of the threshold is not a crucial factor in the construction of an appropriate test.

\subsection{Distinguishing between different sCJD types}

Section \ref{chap:type} confirms that the rfu values cannot be used to distinguish between different subtypes of sCJD, neither on their own or in conjunction with other predictors such as sex and age. Consequently, a test for classifying samples by sCJD subtype is not possible from these data.

\subsection{Limitations and extensions}
Listed here are some of the limitations of this manuscript, and suggestions for possible extensions.
\begin{itemize}
\item This manuscript does not describe the training of SVMs using non-linear kernels, but such SVMs were in fact investigated. Both the Gaussian radial basis function and polynomial kernels \cite{Cristianini00} were tried, but because these kernels cannot be explicitly expressed using a feature mapping, solving the primal solution directly is not possible. Instead the dual must be solved, but due to the inaccuracy of the recovery of the primal solution $(w,b)$ from the dual solution $\alpha$, a reliable cross-validation analysis---which requires $(w,b)$---could not be conducted. However, the unreliable robustness measures on the full set of readings from 0 to 90 hours (with the maximum total fluorescence heuristic invoked) indicated that the Gaussian kernel performed no better than the linear case, with the optimal Gaussian parameter being $\sigma \approx 0.01$, and that the polynomial kernel had a far inferior robustness for polynomial parameter $d>2$. Consequently, linear kernels were deemed sufficient, but this conclusion was not based on any rigorous investigation.
\item Alternative classifying algorithms were not considered. For example logistic regression, a probabilistic binary classifier, has the ability to predict the probability of a sample being sCJD positive given a collection of features, rather than simply assigning it to a class. This essentially interprets the signed distance of a vector to the margin as a probability---zero distance corresponds to an equal probability of being in either class---and so would be able to take advantage of the relationship between maximum fluorescence and age (see Section \ref{sec:classadditional}), allowing the uncertainty of the classification of older patients with low maximum fluorescence to be expressed, rather than immediately classifying these cases as sCJD negative, as the SVM approach has done.

\item  The moving-average smoothing technique used throughout this text is crude. More sophisticated smoothing methods exist, such as the Savitsky-Golay filter \cite{Golay64}, which performs local polynomial regression on a contiguous set of points to find the smoothed point. Consequently, derivative information can be extracted directly from the local polynomials defining each smoothed point by analytic differentiation, and finite differences become unnecessary. Further, ``flattening'' can occur with smoothed curves resulting from a moving-average filters (see Figure~\ref{smoothed} and accompanying discussion) and so first and second derivative approximations will tend to be closer to zero than the actual trend suggests. Savitsky-Golay filters tend not to suffer from this, and so derivative information is more reflective of the trend. However, SVM analysis found no significant meaning in the first and second derivative information, other than their correlation with the maximum fluorescence, and so re-examining this using more sophisticated filters may be fruitless.

\item The noise levels are assumed to be due to accuracy and quality of the equipment and consequently there is no exploration of the possibility of this noise containing information that would aid discrimination between classes. For example, the noise may in fact be oscillatory fluctuations that can reveal something important about the sample, but this is not investigated, and nor are any other possible hypotheses concerning the nature of the noise.

\item By employing the maximum total fluorescence heuristic throughout large parts of this paper, three-quarters of the fluorescence data was ignored (apart from its use in the selection of the replicate itself). It is possible that the relationship between replicates from the same sample could contain useful discriminatory information, but this was not investigated.
\end{itemize}


\bibliographystyle{plain}
\bibliography{biblio}

\end{document}